\documentclass{article}

\usepackage{PRIMEarxiv}

\usepackage{amsmath,amsfonts}
\usepackage{algorithmic}
\usepackage{array}
\usepackage{gensymb}
\usepackage{textcomp}
\usepackage{stfloats}
\usepackage{multirow}
\usepackage{mathtools}
\usepackage[table]{xcolor}
\usepackage{url}
\usepackage{verbatim}
\usepackage[T1]{fontenc}
\usepackage{graphicx}
\usepackage{amsmath}
\usepackage{amssymb}
\usepackage{algorithmic}
\usepackage{caption}
\usepackage{subcaption}
\usepackage[section]{placeins}
\usepackage[ruled,vlined]{algorithm2e}
\newcommand{\R}{\mathbb{R}}
\DeclareMathOperator*{\argminA}{arg\,min}
\DeclareMathOperator*{\argmaxA}{arg\,max}
\def\BibTeX{{\rm B\kern-.05em{\sc i\kern-.025em b}\kern-.08em
    T\kern-.1667em\lower.7ex\hbox{E}\kern-.125emX}}
\usepackage{balance}
\usepackage{hyperref}

\usepackage{booktabs}

\usepackage{physics}
\usepackage{tikz}
\usepackage{tikz-3dplot}
\usepackage[outline]{contour} 
\usepackage{xcolor}

\newcommand{\todo}[1]{\textcolor{red}{#1}}

\newcommand{\alk}{\breve{\pi}} 
\newcommand{\XY}{XY} 

\colorlet{veccol}{green!50!black}
\colorlet{projcol}{blue!70!black}
\colorlet{myblue}{blue!80!black}
\colorlet{myred}{red!90!black}
\colorlet{mydarkblue}{blue!50!black}
\tikzset{>=latex} 
\tikzstyle{proj}=[projcol!80,line width=0.08] 
\tikzstyle{area}=[draw=veccol,fill=veccol!80,fill opacity=0.6]
\tikzstyle{vector}=[-stealth,myblue,thick,line cap=round]
\tikzstyle{unit vector}=[->,veccol,thick,line cap=round]
\tikzstyle{dark unit vector}=[unit vector,veccol!70!black]
\usetikzlibrary{angles,quotes} 
\contourlength{1.3pt}

\title{Fast and robust single particle reconstruction in 3D fluorescence microscopy
}

\author{
  Thibaut Eloy$^1$, Etienne Baudrier$^1$, Marine Laporte$^2$, Virginie Hamel$^2$, Paul Guichard$^2$, Denis Fortun$^1$ \\[5pt]
  $^1$ ICube - UMR7357, CNRS, University of Strasbourg, France \\
  $^2$ Centriole Lab, Department of Molecular and Cellular Biology, University of Geneva, Geneva, Switzerland\\
}

\begin{document}
\maketitle

\begin{abstract}
Single particle reconstruction has recently emerged in 3D fluorescence microscopy as a powerful technique to improve the axial resolution and the degree of fluorescent labeling. It is based on the reconstruction of an average volume of a biological particle from the acquisition multiple views with unknown poses. Current methods are limited either by template bias, restriction to 2D data, high computational cost or a lack of robustness to low fluorescent labeling. In this work, we propose a single particle reconstruction method dedicated to convolutional models in 3D fluorescence microscopy that overcome these issues. We address the joint reconstruction and estimation of the poses of the particles, which translates into a challenging non-convex optimization problem. Our approach is based on a multilevel reformulation of this problem, and the development of efficient optimization techniques at each level. We demonstrate on  synthetic data that our method outperforms the standard  approaches in terms of resolution and reconstruction error, while achieving a low computational cost. We also perform	 successful reconstruction on real datasets of centrioles to show the potential of our method in concrete applications.
\end{abstract}

\section{Introduction}

Fluorescence microscopy has experienced great resolution improvements in recent years thanks to progress in super resolution microscopy \cite{Schermelleh19} and more recently in expansion microscopy \cite{Gambarotto19}. These advances have motivated the use of fluorescence microscopy in structural biology, to decipher the structure of macromolecular assemblies  that were previously observable only with electron microscopy \cite{Chen15}. However, it is still limited by two issues. Firstly, the resolution of most fluorescence microscopes is strongly anisotropic (the resolution in the axial direction of the microscope is typically 3 to 5 times lower than in the lateral plane). Secondly, spurious intensity variations along the protein structures  can hinder the biological interpretation. This phenomenon is explained by a low and non uniform degree of labeling (DOL), which characterizes the density of fluorescent dyes that attach to the protein of interest and emit fluorescent light.

We address these issues with a single particle reconstruction (SPR) approach.  The principle is to combine several acquisitions of identical biological particles observed from different orientations, to reconstruct a single particle model. Thanks to the fusion of complementary information in each view, we aim at reconstructing volumes with isotropic resolution and uniform DOL. This approach has received an increasing attention in recent years and offers new perspectives in structural biology to decipher the protein architecture of large macromolecular assemblies \cite{Salas17,Sieben18,Gambarotto19,Mahecic20,Heydarian21,Mendes22,Liu22}.

In this work, we focus on a category of fluorescence microscopy techniques that we call \textit{convolutional modalities}, for which the observation model is the convolution with a point spread function (PSF). Confocal and STED microscopy \cite{Schermelleh10} are the main representatives of this category and have been proven to be well suited to observe important macromolecular assemblies such as the centriole or the nuclear pore, in particular when combined with expansion microscopy \cite{Gambarotto19,Mahecic20}.

\subsection{Challenges and related works}
\label{sec:related_work}

Multiview reconstruction is a common practice in several microscopy systems for which physical calibration procedures provide an accurate estimation of the pose of each view. (e.g. selective plane illumination microscopy,  \cite{Temerinac12,Preibisch14}, 3D total-internal reflection fluorescence microscopy \cite{Boulanger14}). By contrast, the specificity  of the multiview reconstruction problem in SPR is the absence of prior knowledge about the poses. This results in a particularly challenging non-convex optimization problem that requires the development of dedicated methods. 
 
SPR has a longstanding and successful history in another modality called cryo-electron microscopy (cryo-EM), for which a lot of efficient reconstruction methods have been proposed over the past 20 years \cite{Carazo15,Punjani16}. However, the imaging model of cryo-EM fundamentally differs from  the one of convolutional modalities. The cryo-EM images are 2D projections of the 3D densities, whereas convolutional modalities acquire 3D images by convolution of the fluorescent signal with a PSF. Despite this difference, the most common practice for SPR in fluorescence imaging is to apply cryo-EM reconstruction methods on 2D fluorescence data \cite{Salas17,Sieben18}, which ignores the true imaging model of optical microscopes and leads to coarse and suboptimal reconstructions. 

Most methods developed specifically for fluorescence microscopy are either restricted to 2D input images \cite{Loschberger12,Szymborska13,Blundell21}, or require to use an initial template or a parametric model \cite{Mennella12,Broeken15,Shi19}, which inevitably introduces a bias in the reconstruction. The method we proposed in \cite{Fortun18} is the only one, up to our knowledge, to take into account an appropriate 3D imaging model. Despite  good performances on real data \cite{Gambarotto19,Mahecic20}, it has some major practical limitations. Firstly, it requires a classification step to reduce the number of particles. Automatic classification often lack robustness on real data, and manual particle selection is tedious and introduces an arbitrary bias. The computation time of \cite{Fortun18} can also be prohibitive for non-symmetric particles, for which the space of the registration parameters cannot be reduced. Finally, it is based on a sequential optimization procedure that considers the views one by one in a predefined order, which leads to reconstruction errors for low fluorescent labeling rates.

The SPR principle has also been applied in
single molecule localization microscopy (SMLM), a fluorescence imaging  technique that generates data in the form of point clouds. A few works have been dedicated to SPR for SMLM data from the point of view of point clouds registration \cite{Broeken15,Heydarian21}. Our reconstruction problem with convolutional modalities is radically different since our data is a collection of volumes, which cannot be processed with the same methods as for point clouds. Moreover, our goal is reconstruction and not only registration.

\subsection{Contributions}
We propose a new method to estimate jointly the poses (rotation and translation) of the particles and the reconstruction in SPR for convolutional modalities in fluorescence imaging. Our approach is based on a reformulation of the original joint optimization problem with respect to (w.r.t.) the volume, the rotations and the translations, into an equivalent hierarchical optimization scheme that decouples the minimization w.r.t. each variable in three nested levels. The volume is estimated at the highest level with stochastic optimization, and the key for the success of our approach is the development of fast approximate solvers for the minimization w.r.t. to the rotations and translations at the lower levels. Thanks to this approach, the computational cost of a reconstruction is reduced to less than 30 minutes, with a reference-free 
initialization. Moreover, our approach does not require any critical hyper-parameter tuning, which simplifies its usage in practice. We demonstrate on several synthetic datasets that we  outperform the standard methods in terms of resolution and visual assessment. Finally, we show the potential of our approach on real data of centrioles acquired with a combination of confocal and expansion microscopy. 


Our work builds upon the cryo-EM reconstruction method cryoSPARC described in \cite{Punjani16}. We deviate from it on the main following points:
\begin{itemize}
\item cryoSPARC cannot be applied to our data since it is designed for a forward model of 2D projection. Instead, we develop a method dedicated to 3D convolutional models.
\item We estimate the poses with a multilevel optimization scheme, whereas the strategy of \cite{Punjani16} is to reconstruct the volume without pose estimation by maximizing a marginalized likelihood.
\item We estimate the translations at the last level of our hierarchical scheme with a fast phase correlation algorithm. The decoupling of translations and rotations estimation is crucial for computational tractability in our 3D case.
\end{itemize}


The organization of the manuscript is as follows. In Section \ref{part2}, we present the observation model and formulate the reconstruction as an optimization problem. In Section \ref{part3}, we present our optimization method. In Section \ref{part4} we present the experimental results obtained both on simulated and real datasets.

\section{Problem formulation}

\label{part2}

\subsection{Background on the representation of  rotations}
\label{sec:background}

\begin{figure}
\centering
\tdplotsetmaincoords{60}{110}
\begin{tikzpicture}[scale=2,tdplot_main_coords]
  
  \def\rvec{1}
  \def\thetavec{40}
  \def\phivec{60}
  
  \coordinate (O) at (0,0,0);
  \draw[thick,->] (0,0,0) -- (1,0,0) node[below left=-3]{$x$};
  \draw[thick,->] (0,0,0) -- (0,1,0) node[right=-1]{$y$};
  \draw[thick,->] (0,0,0) -- (0,0,1) node[above=-1]{$z$};
  
  \tdplotsetcoord{P}{\rvec}{\thetavec}{\phivec}
  \draw[vector,red] (O)  -- (P) node[above right=-2] {d};
  \draw[dashed,myred]   (O)  -- (Pxy);
  \draw[dashed,myred]   (P)  -- (Pxy);
  
  \tdplotdrawarc[->]{(O)}{0.2}{0}{\phivec}
    {anchor=north}{$\phi_1$}
  \tdplotsetthetaplanecoords{\phivec}
  \tdplotdrawarc[->,tdplot_rotated_coords]{(0,0,0)}{0.4}{0}{\thetavec}
    {anchor=south west}{\hspace{-1mm}$\phi_2$}

\end{tikzpicture}
\caption{Representation of a direction $d$ (point on the unit sphere) with its azimuth $\phi_1$ and inclination $\phi_2$.}
\label{spherical coordinates}
\end{figure}
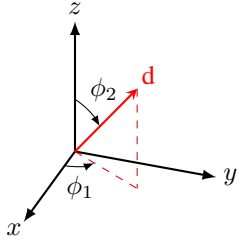

We use the axis-angle representation $(d,\psi)$ to represent a 3D rotation, where $d \in \R ^ 3$ is the rotation axis, and $\psi \in [0,2\pi[$ is the rotation angle around this axis. The  axis $d$ is the unit vector:
\begin{equation}
d = 
\begin{pmatrix}
\cos(\phi_1)\sin(\phi_2) \\
\sin(\phi_1)\sin(\phi_2) \\
\cos(\phi_2) \\
\end{pmatrix} ,
\end{equation}
where $\phi_1 \in [0, 2\pi[$ is the azimuth and $\phi_2 \in [0, \pi[$ is the inclination (see Figure \ref{spherical coordinates}). 

Our method requires a discretization of SO(3). We write it $\{\theta_{i,j}=(\phi_{1,j},\phi_{2,j},\psi_i)\,|\; (j,i) \in [\![0, M_{d}-1]\!]\times [\![0, M_\psi-1]\!] \}$, where $M_d$ and $M_{\psi}$ are the discretization sample sizes 
of $(\phi_{1},\phi_{2})$ and $\psi$, respectively. For $\psi$, we choose a uniform discretization of $[0, 2\pi[$: $\forall i \in [\![0,M_{\psi}-1]\!]:\:\ \psi_i = 2\pi i/M_{\psi}$. For $(\phi_{1},\phi_{2})$, we use the Fibonacci discretization, which gives an almost uniform discretization of the unit sphere. It is formalized as follows 
\begin{align}
    \forall j \in [\![0,M_d-1]\!],\quad \phi_{1,j} \equiv \frac{2\pi j}{G} [2\pi] \\
    \phi_{2,j} = \arccos(1 - \frac{2j+1}{M_{d}})\enspace,
\end{align}
with $G = \frac{1+\sqrt{5}}{2}$ the golden ratio.

\subsection{Forward model} \label{obs_model_sec}

Let us denote $\{y_l\}_{l \in [\![1,N]\!]}$ a set of $N$ volumes (we call volume a function $\Omega\to \R$ where $\Omega$ is a bounded domain in $\R^3$), where $y_l$ is a view that	 contains a single particle with an orientation $\theta_l \in \R ^ 3$ and a translation $t_l \in \R ^ 3$. From this set of views, the goal is to reconstruct a volume  $f$, that represents the observed particle. The image acquisition model belongs to convolutional modalities, defined by 
\begin{equation}
    y_l(x) = (h*\mathcal{T}_{t_l}(\mathcal{R}_{\theta_l}(f)))(x) + \varepsilon(x) \label{mod_obs_eq},
\end{equation}
where $h$ is the PSF of the microscope, $\varepsilon$ is an additive Gaussian noise, $\mathcal{T}_{t}$ is the translation operator of vector $t$ defined by
\begin{equation}
	 \mathcal{T}_{t}(f)(x) = f(x-t),
\end{equation}
and $\mathcal{R}_{\theta}$ is the rotation operator of angle $\theta$  defined by 
\begin{equation}
	\mathcal{R}_{\theta}(f)(x) = f(R_{\theta}^Tx),
\end{equation}
where $R_{\theta} \in SO_3(\R)$ is the rotation matrix of angle $\theta$.

We do not make any assumption on the form of the PSF, though in practice it has an anisotropic shape, more elongated along the Z axis \cite{kirshner20133}, which creates the anisotropy of resolution.

\subsection{Joint minimization problem}

We formulate the reconstruction as a maximum likelihood estimation of the volume, the rotation and the translation parameters, which amounts to the following least squares problem:
\begin{equation}
f^ *, \Theta ^* , T^ * = \argminA_{f, \Theta, T}  E(f, \Theta, T) , \label{main_prob}
\end{equation}
where
\begin{equation}
    E(f, \Theta, T) = \sum_{l=1}^{N} \|y_l - h*\mathcal{T}_{t_l}(\mathcal{R}_{\theta_l}(f))\|_2^2 			\label{energie} ,
\end{equation}
with
$$\Theta = \{\theta_l\}_{l \in [\![1,N]\!]} \text{ and } T=\{t_l\}_{l \in [\![1,N]\!]}.$$ A regularization term could be added in \eqref{energie}, but we found  experimentally that it was not necessary to obtain satisfying results, thanks to the averaging effect over several particles.

To reduce the computational cost, we formulate the energy~\eqref{energie} in the Fourier domain. It allows us to perform convolutions with simple pointwise multiplication. Besides, rotation and Fourier transform commute, and the translation in the Fourier domain is simply a phase shift. Formally, we have 
\begin{equation}
    \mathcal{F}(h*\mathcal{T}_{t_l}(\mathcal{R}_{\theta_l}(f))) = 
     \hat h \rho_{t_l} \mathcal{R}_{\theta_l}(\hat f),
\end{equation}
where $\mathcal{F}$ is the Fourier transform, $\rho_{t_l}$ is a phase factor defined by $\rho_{t_l}: \omega \mapsto e^{it_l\omega}$, and we use the notation $\mathcal{F}(f) = \hat f$. The energy, written in the Fourier domain is obtained with Parseval's theorem: 
\begin{equation}
    E(f, \Theta, T) = \sum_{l=1}^{N} \|\hat y_l - \hat h \rho_{t_l}  \mathcal{R}_{\theta_l}(\hat{f})  \|_2^2. \label{energy_fourier}
\end{equation}

\section{Optimization}
\label{part3}
The optimization problem \eqref{main_prob} is non-convex with many local minima. 
It is reminiscent of auto-calibration problems in microscopy, where the volume and some parameters of the observation model are estimated jointly. The alternating optimization scheme commonly used for auto-calibration is unable to escape from local minima, and it relies on theoretical observation models (such as PSF models in blind deconvolution) that provide a good initialization of the parameters. In our case, an alternating scheme is destined to fail because we do not have any initial guess of the volume or the poses.

\subsection{Multilevel reformulation of the optimization problem}

To overcome this issue, we reformulate the problem \eqref{main_prob} in the multilevel equivalent form 
\begin{subequations}
\begin{align}
    {\hat f}^{*} = \argminA_{\hat f} E(f, \Theta^{*}, T^{*}) \label{main} \\  
    \textrm{s.t.} \:\: \Theta^{*} = \argminA_{\Theta} E(f, \Theta, T^{*}) \label{contrainte} \\
    \textrm{s.t.}  \:\: T^{*} = \argminA_T E(f, \Theta, T).\label{trans}
\end{align}
\label{reform}
\end{subequations}
In this scheme, a subproblem at a given level in \eqref{reform} is nested as a constraint in the subproblem of the upper level. 
 The main interest of this reformulation is to decompose the original intractable problem into three tractable subproblems, similarly with alternating optimization. However, differently from an alternating scheme, the locally optimal points of the joint problem \eqref{main_prob} are also optimal points of the multilevel formulation. 
The counterpart is that if an iterative method is used to solve a problem at a given level,  the lower level problem must be solved at each iteration. Thus, the computational cost of this nested optimization  can become prohibitive if the solvers of the subproblems are not efficient enough (for more details on multilevel optimization see \cite{sato2021gradient}). In the following three subsections, we detail our fast and complementary optimization strategies for the subproblems in~\ref{reform}.

\subsection{Optimization with respect to the volume}
\label{part:opt_volume}
When the pose parameters $\Theta$ and $T$ are known, the minimization problem \eqref{main} can be solved easily in closed form~\cite{Fortun18}. However, since pose estimation is formulated as constraints \eqref{contrainte} and \eqref{trans}, we have to adopt an iterative approach to update the poses at each iteration. Moreover the constrained problem is non-convex, and the sum over a potentially large number $N$ of particles can dramatically increase the computational cost. To overcome these issues, we use a simple stochastic gradient descent strategy. The  volume is updated with the following equation: 
\begin{equation}
\label{maj}
    {\hat f}^{(n+1)}(\omega) = {\hat f}^{(n)}(\omega) - \mu  \nabla_{\hat f} E_{l_n}({f }^{(n)}, {\theta^*_{l_n	}}, {t^*_{l_n}})(\omega),
\end{equation}
where $l_n\in[\![1,N]\!]$ is a randomly selected view index at iteration $n$, $E_l(f, {\theta_{l}}, {t_{l}}) = \|\hat y_{l} - \hat h  \rho_{t_{l}} \mathcal{R}_{\theta_{l}}({\hat f })  \|_2^2 $ is the energy associated to index $l$ and $\mu$ is the gradient descent step. The random indices $l_n$ are imposed to cycle through all the views, in order to divide the optimization procedure in epochs. The computational cost at iteration $n$ is reduced to the evaluation of the the gradient of $E_{l_n}$ (the derivation of the gradient is detailed in Appendix \ref{grad_comp}).
 The robustness of SGD to non-convex problems has been established for similar objective functions in the context of learning applications or for cryo-EM reconstruction \cite{Punjani16}, though convergence properties can be guaranteed only in the convex case.

\subsection{Optimization with respect to the orientations}

The secondary optimization problem \eqref{contrainte}, dedicated to the estimation of the orientations, has to be solved at each iteration of the main problem \eqref{main} for the evaluation of the gradient $\nabla_{\hat f} E_{l_n}$.
The $N$ views are considered independently, so \eqref{contrainte} can be rewritten as: 
\begin{equation}
\forall l \in [\![1,N]\!] \:\: \theta_l ^ {*} = \argminA_{\theta_l} E_l(f, \theta_l, t_l ^ {*}). \label{prob_or}
\end{equation}
A gradient based approach to solve \eqref{prob_or} is very likely to reach a poor local minimum because of the non convexity of the problem, such that it can be used only for refinement. On the other hand, an exhaustive search over a discretization of $SO(3)$ would be computationally too expensive.

To tackle this issue, we perform an exhaustive search on a restricted set of orientations $\{\theta_{i,j} |\, (j,i)\in \mathcal{I}^{\psi,l}\times \mathcal{I}^{d,l}\}$, where $\mathcal{I}^ {\psi, l} \subset [\![0, M_{\psi}-1]\!]$ and $\mathcal{I}^{d,l} \subset [\![0, M_{d}-1]\!]$ are subsets of indices in the SO(3) discretization defined in Section \ref{sec:background}. The sizes $N_d$ and $N_{\psi}$ of $\mathcal{I}^{l,d}$ and $\mathcal{I}^{l,\psi}$, respectively, are chosen such that $N_d \ll M_d$ and $N_{\psi} \ll M_{\psi}$. Let us define $E^ l_{i,j} = E_l(f, \theta_{i,j}, t_l^ {*})$ the energy associated to orientation $\theta_{i,j}$. The solution of \eqref{prob_or} is approximated by $\theta_{i^ *,j^ *}$ with $$i ^ *,  j ^ * = \argminA_{i \in \mathcal{I}^{\psi,l}, j \in \mathcal{I}^{d,l}} E^l_{i,j}.$$ 

The critical issue is to design a fast   method to build subsets $\mathcal{I}^ {\psi, l}$ and $\mathcal{I}^ {d, l}$ that are adapted to the landscape of the energy $E^l_{i,j}$. Let us define $\mathcal{Q}^{l, d}$ and $\mathcal{Q}^{\psi, d}$ the sampling distributions from which we create $\mathcal{I}^ {\psi, l}$ and $\mathcal{I}^ {d, l}$. Since we are looking for the maximum likelihood in \eqref{prob_or}, we can follow the approach of \cite{Punjani16} and define $\mathcal{Q}^{l, d}$ and $\mathcal{Q}^{\psi, d}$ from 
the marginal likelihoods w.r.t $d$ and $\psi$, respectively. These marginal likelihoods are defined at each point of the discretization by
\begin{align}
\forall i\in[\![0, M_{d}-1]\!], \: \: \pi^{l,\psi}_i = \sum_{j=0}^{M_d-1} p^l_{i,j} \label{eq:marg1},\\
\forall j\in[\![0, M_{\psi}-1]\!], \: \: \pi^{l,d}_j = \sum_{i=0}^{M_{\psi}-1} p^l_{i,j} \label{eq:marg2},
\end{align}
where $p_{i,j} ^ l \propto \exp(-E^l_{i,j})$ is the likelihood associated to the orientation $\theta_{i,j}$. 
However the likelihoods are unknown and we need to approximate them. The key observation is that the variation of the volume $\hat f$ between two consecutive iterations of \eqref{maj} is small. Therefore, the solution of \eqref{prob_or} at a given iteration is close to the solution at the previous iteration. Thus, we can reasonably approximate the likelihoods in \eqref{eq:marg1} and \eqref{eq:marg2} by their values at the previous iterations (in what follows, $p^l_{i,j}$ refers to its value at the previous iteration, without notation change for the sake of readability).

The difficulty is that the likelihood is not evaluated for all the elements of the sums in \eqref{eq:marg1} and \eqref{eq:marg2}, but only at the indices $\mathcal I^{l,d}$ and $\mathcal I^{l,\psi}$. This issue is overcome by using importance sampling \cite{Tokdar} to approximate the marginal likelihoods \eqref{eq:marg1} and \eqref{eq:marg2}. Since the indices are drawn from the  distributions $\mathcal{Q}^{l, d}$ and $\mathcal{Q}^{\psi, d}$, they can be used as importance distributions to create the following approximations of $\pi^ {l, \psi}$ and $\pi ^ {l,d}$: 
\begin{align}
     \forall i \in \mathcal{I}^{l,\psi} \: \: \tilde{\pi}^{l,\psi}_i = \sum_{j \in \mathcal{I}^{l,d}} \frac{p^l_{i,j}}{\mathcal{Q}^{l,d}_j} ,\\
    \forall j \in \mathcal{I}^{l,d} \: \: \tilde{\pi}^{l,d}_j = \sum_{i \in \mathcal{I}^{l,\psi}} \frac{p^l_{i,j}}{\mathcal{Q}^{l,\psi}_i}.
\end{align} 
To compute the approximated likelihoods $\alk^{l,\psi}$ and $\alk^{l,d}$ associated to all the discretization elements of SO(3), a kernel estimation method is used:
\begin{align}
        \forall i \in [\![0, M_{\psi}-1]\!], \: \: \alk_{i}^{l,\psi} =  Z_{\psi}^{-1} \sum_{k \in \mathcal{I}^{l,\psi}} K_{i, k}^{\psi} \tilde{\pi}^{l,\psi}_k , \\
        \forall j \in [\![0, M_d-1]\!], \: \: \alk_{j}^{l,d} = Z_d^{-1} \sum_{k \in \mathcal{I}^{l,d}} K_{j, k}^{d} \tilde{\pi}^{l,d}_k ,
\end{align}
where $Z_{\psi}$ and  $Z_d$ are normalization constants given by 
$$
 Z_{\psi} = \sum_{i=1}^{M_{\psi}} \sum_{k \in \mathcal{I}^{l,\psi}} K_{i, k}^{\psi} \tilde \pi^{l,\psi}_k
 \text{ and }
 Z_d = \sum_{j=1}^{M_d} \sum_{k \in \mathcal{I}^{l,d}} K_{j, k}^d \tilde \pi^{l,d}_k ,
$$
and $K^{\psi}$ and $K^d$ are kernels defined by: 
\begin{equation}
\forall i,k \in [\![0, M_{\psi}-1]\!],\: \: K^{\psi}_{i,k} = \exp(\beta_{\psi}\cos(\psi_i - \psi_k)), \label{kernel2}
\end{equation}
\begin{equation}
\forall j,k \in [\![0, M_{d}-1]\!], \: \: K^{d}_{j,k} = \exp(\beta_d d_{j}^{T}d_{k}), \label{kernel1}
\end{equation}
where $\beta_{\psi}$ and $\beta_d$ are hyperparameters.

In the first iterations of \eqref{maj}, the hypothesis of small variation can be violated until reaching a stable coarse reconstruction. To account for this, the final sampling distributions are defined as linear combinations of the approximated likelihoods with a uniform distribution:
\begin{align}
    \mathcal{Q}^{l,d} = \alpha \mathcal{U}^d + (1-\alpha) \alk^{l,d} , \label{Q}\\
    \mathcal{Q}^{l,\psi} = \alpha \mathcal{U}^{\psi} + (1-\alpha) \alk^{l,\psi}, 
\end{align}
where  $\mathcal{U}^{\psi} = \frac{1}{M_{\psi}}$ and $\mathcal{U}^d = \frac{1}{M_d}$ are the uniform components, and $\alpha$ is the coefficient adjusting the proportion of uniform component. The uniform component enables to explore the entire rotation space at the beginning of the reconstruction process, and the evolution of $\alpha$ gradually restricts the search space around plausible solutions. It is set to 1 at the beginning of the gradient descent, and is divided by $\alpha_r$ between two epochs of the SGD \eqref{maj}.

\subsection{Optimization with respect to the translations}

The problem \eqref{trans} has to be solved at each evaluation of the energy $E$ in the minimization of \eqref{contrainte} described in the previous section. It can be written
\begin{equation}
\forall l \in [\![1, N]\!],~ t_l^{*} = \argminA_{t_l}\|\hat{y_l} - \rho_{t_l}\hat b_l\|_2^2, \label{eq:phase_corr}
\end{equation}
where $\hat{b}_l = \hat{h}\mathcal{R}_{\theta_l}(\hat{f})$ is the rotated and convolved  volume in the Fourier domain. The problem amounts to the estimation of a phase shift $\rho_{t_l}$ of $\hat{b}_l$ to match the view $\hat{y_l}$. It can be straightforwardly reformulated as the maximization of the cross-correlation \cite{Fienup97}:
\begin{equation}
 t_l^{*} = \argmaxA_{t_l}  \int\rho_{t_l}(\omega)\,\hat{y_l}(\omega)\, \overline{\hat{b}_l}(\omega)d\omega
 , \label{eq:phase_corr_reform}
\end{equation}
where  $\bar{z}$ refers to the conjugate of $z$.
Since the phase shift corresponds to a translation in the spatial domain, the most efficient way to solve \eqref{eq:phase_corr_reform} is to find the spatial position of the correlation peak in the inverse Fourier transform of the cross-correlation. In practice we use a standard phase correlation algorithm which considers a normalized cross-correlation instead of a simple cross-correlation in \eqref{eq:phase_corr_reform}, in order to get a sharper and more accurate correlation peak \cite{Tong19}. The computational cost of this step  essentially lies in a single inverse Fourier transform. This efficiency is crucial for the computational tractability of the overall reconstruction algorithm since this step is nested in the two upper problems \eqref{main} and \eqref{contrainte}. 

The detailed reconstruction steps are regrouped in Algorithm~\ref{algo_summary}.

\begin{algorithm}[t]
\caption{Reconstruction algorithm }
\begin{algorithmic}[1] 
\STATE Initialize $\hat f$ randomly
\STATE Initialize probability distributions, with uniform distributions
    $\mathcal{Q}^{l,\psi} = \mathcal{U}$ and $\mathcal{Q}^{l,d} = \mathcal{U}$
\WHILE{stopping criteria} \label{main_loop}
\STATE Randomly mix the views
\FOR{$l$ =1 to $N$}
\STATE Randomly draw indices $\mathcal{I}^{l, \psi}$ and $\mathcal{I}^{l,d}$ from $\mathcal{Q}^{l,\psi}$ and $\mathcal{Q}^{l,d}$
\FOR{$(i,j) \in \mathcal{I}^{l,\psi} \times \mathcal{I}^{l,d}$} \label{begin_est_rot}
\STATE $\theta_{i,j} = (d_j, \psi_i)$
\STATE Compute $\hat h\mathcal{R}_{\theta_{i,j}}(\hat f)$ \label{rot}
\STATE Evaluate $t_l^{*}(\theta_{i,j})$, using phase correlation between $\hat h\mathcal{R}_{\theta_{i,j}}(\hat f)$ and $\hat y_l$. \label{trans_est}
\STATE Compute $E^l_{i,j} =  \|\hat h\rho_{t_l^*(\theta_{i,j})}\mathcal{R}_{\theta_{i,j}}(\hat f) - \hat y_l\|_2^2$  \label{energy_calc}
\ENDFOR 
\STATE 
$\theta_l^{*} = (d_{j^*}, \psi_{i^*})$ with $i^*, j^* = \argminA
\{ E^l_{i,j} \,|\, i,j \in \mathcal{I}^{l,\psi}\times\mathcal{I}^{l,d}\}$ \label{end_est_rot}
\STATE $\forall (i,j) \in \mathcal{I}^{l, \psi} \times \mathcal{I}^{l, d}$,
 compute $p^l_{i,j}=\exp(-E^l_{i,j})$.
\STATE Compute marginalized likelihoods:
\STATE $\forall i \in \mathcal{I}^{l,\psi} \: \: \tilde{\pi} ^{l,\psi}_i = \sum_{j \in \mathcal{I}^{l,d}} \frac{p^l_{i,j}}{\mathcal{Q}^{l,d}_j}$
\STATE $\forall j \in \mathcal{I}^{l,d} \: \: \tilde{\pi}^{l,d}_j = \sum_{i \in \mathcal{I}^{l,\psi}} \frac{p^l_{i,j}}{\mathcal{Q}^{l,\psi}_i}$
\STATE Update probability distributions 
\STATE $\mathcal{Q}^{l,d}_j = \alpha \mathcal{U}^d_j + (1-\alpha) Z_{d}^{-1} \sum_{k \in \mathcal{I}^{l,d}} K_{j, k}^{d} \tilde{\pi}^{l,d}_k$
\STATE $\mathcal{Q}^{l,\psi}_i = \alpha \mathcal{U}^{\psi}_i + (1-\alpha) Z_{\psi}^{-1} \sum_{k \in \mathcal{I}^{l,\psi}} K_{i, k}^{\psi} \tilde{\pi}^{l,\psi}_k$
\STATE $\alpha = \alpha / \alpha_{r}$
\STATE  $\hat f =  \hat f - \mu \nabla_{\hat f} E_l(f,\theta_l^{*},t_l^{*})$ \label{update_eq}
\ENDFOR
\ENDWHILE
\STATE ${\hat f}^* = \hat f$ 
\RETURN $\mathcal{F} ^ {-1}({\hat f}^*), (\theta_l^{*})_{l \in [\![1, N]\!]}, (t_l^{*})_{l \in [\![1, N]\!]}$
\end{algorithmic}
\label{algo_summary}
\end{algorithm}

\section{Experimental results}
\label{part4}
In this section, we present experiments to evaluate the performance of our method. First, we show reconstructions on synthetic data sets and compare our results with other standard approaches. We also analyze the robustness of our method to different imaging conditions and its sensitivity to hyperparameters. Finally, we show results obtained on a real dataset of centrioles.

\subsection{Simulated data}

\begin{figure}
\centering
\begin{tabular}{m{10pt}@{\hspace{2pt}}m{50pt}@{\hspace{2pt}}m{50pt}@{\hspace{10pt}}m{50pt}@{\hspace{2pt}}m{50pt}@{\hspace{10pt}}m{50pt}@{\hspace{2pt}}m{50pt}@{\hspace{10pt}}m{50pt}@{\hspace{2pt}}m{50pt}}
&\multicolumn{2}{c}{\centering Clathrine} & \multicolumn{2}{c}{\centering NLR Resistosome} & \multicolumn{2}{c}{\centering HIV Vaccine} & \multicolumn{2}{c}{\centering AMPA receptors} \\
\multirow{2}{*}{\rotatebox[origin=c]{90}{Ground truth~~~~}}&
\includegraphics[width=50pt]{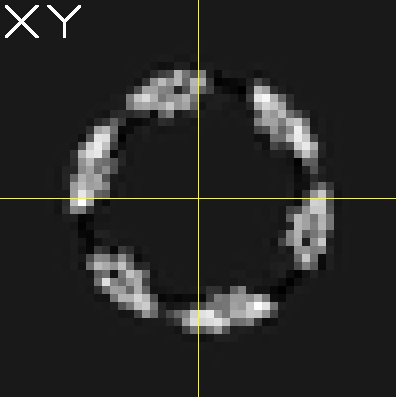} &
\includegraphics[width=50pt]{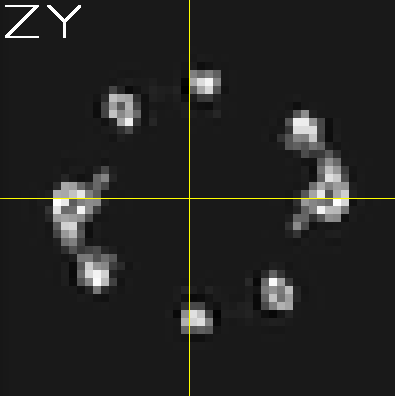} &
\includegraphics[width=50pt]{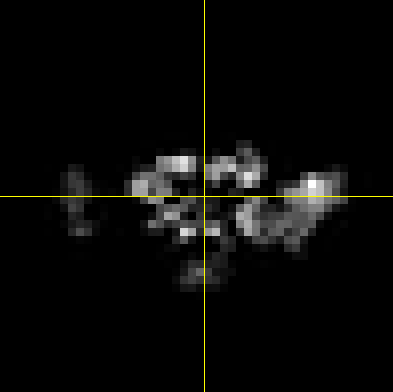} &
\includegraphics[width=50pt]{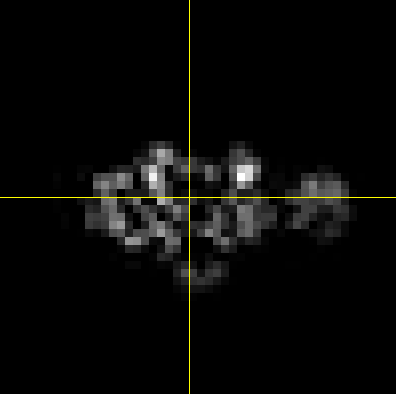} & 
\includegraphics[width=50pt]{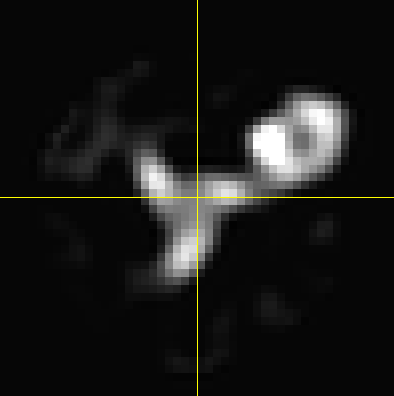} &
\includegraphics[width=50pt]{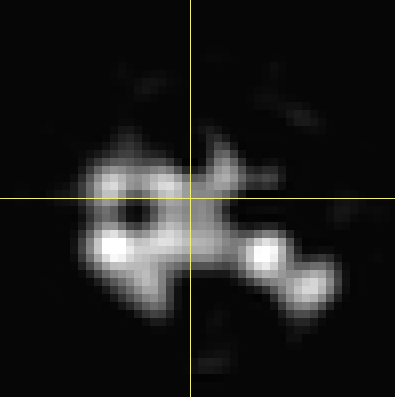} &
\includegraphics[width=50pt]{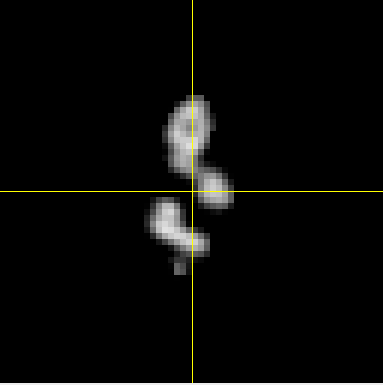} &
\includegraphics[width=50pt]{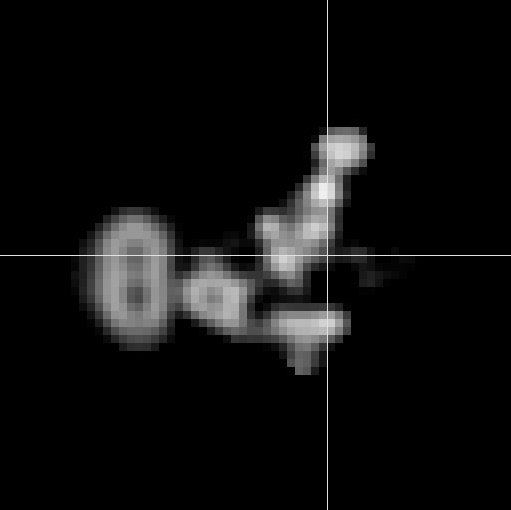}
\\[-2pt]
&\includegraphics[width=50pt]{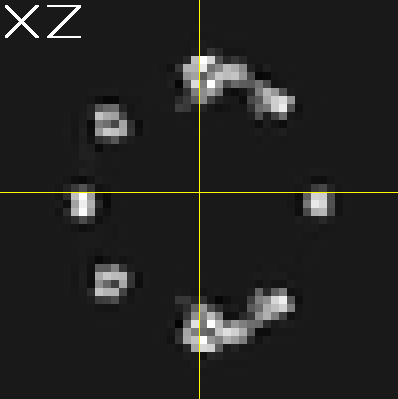} &
\includegraphics[width=50pt]{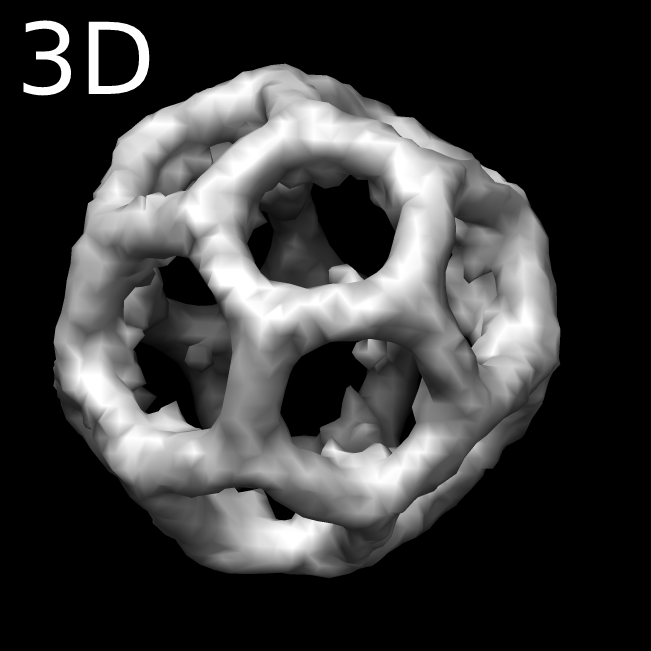} &
\includegraphics[width=50pt]{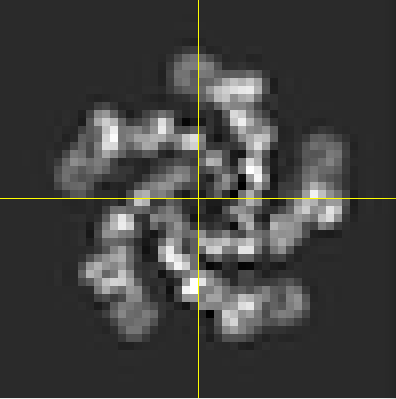} &
\includegraphics[width=50pt]{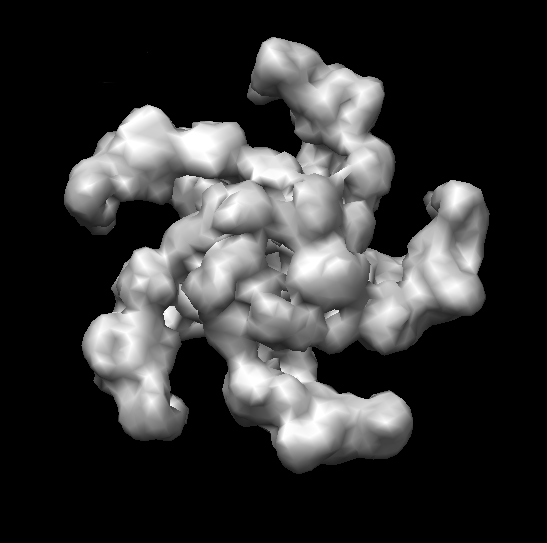} &
\includegraphics[width=50pt]{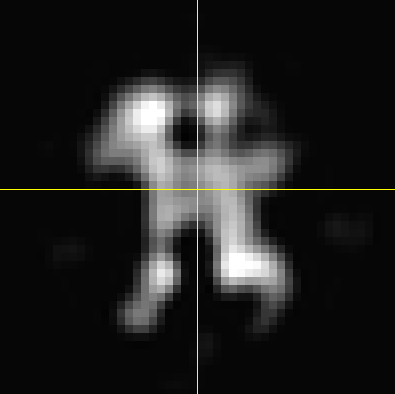} &
\includegraphics[width=50pt]{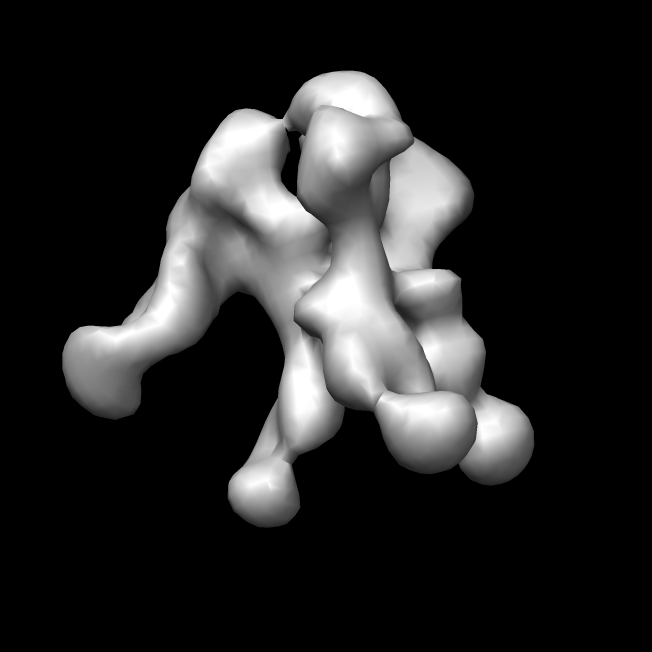} &
\includegraphics[width=50pt]{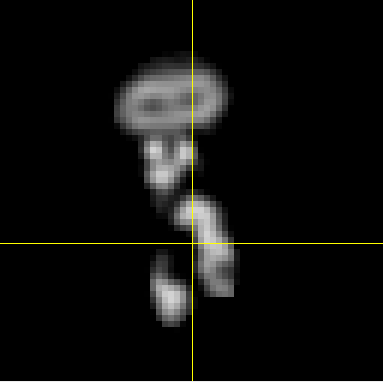} &
\includegraphics[width=50pt]{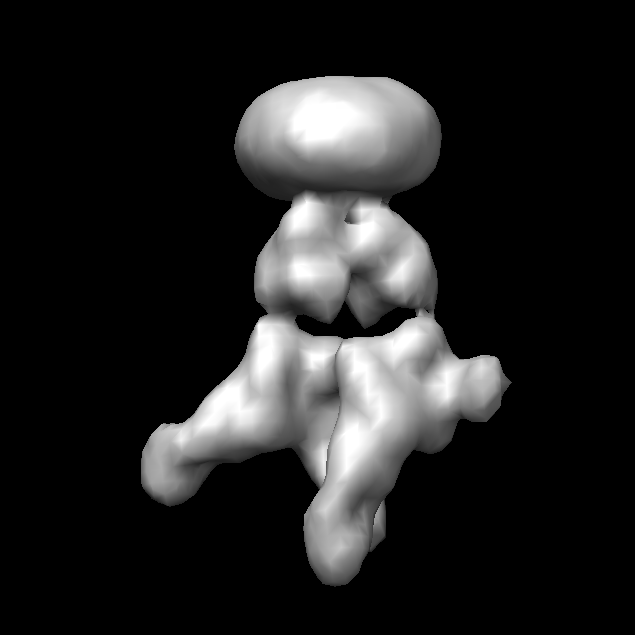} 
\\[25pt]
\multirow{2}{*}{\centering \rotatebox[origin=c]{90}{Example of view}}&
\includegraphics[width=50pt]{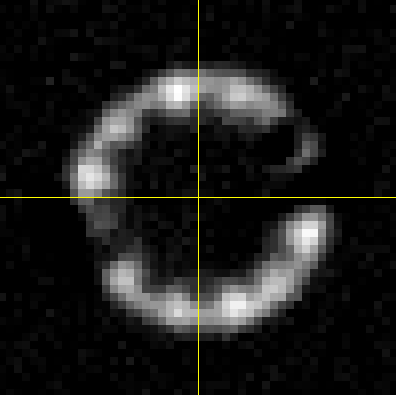} &
\includegraphics[width=50pt]{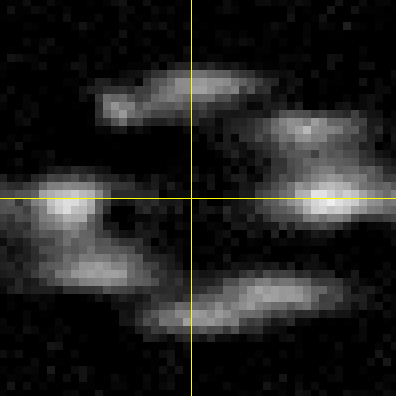} & 
\includegraphics[width=50pt]{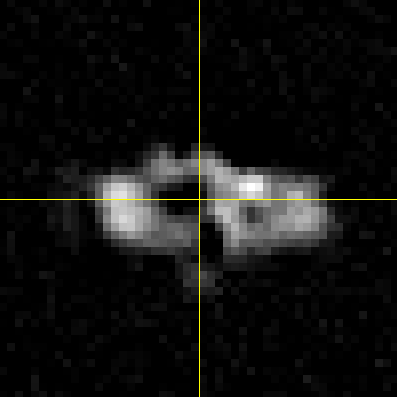} &
\includegraphics[width=50pt]{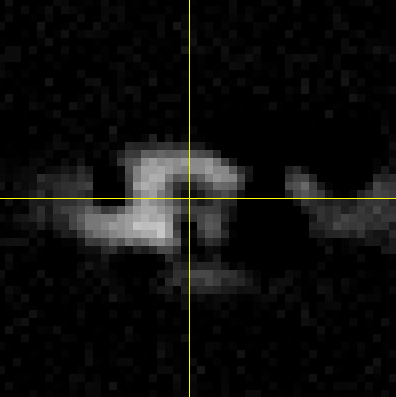} &
\includegraphics[width=50pt]{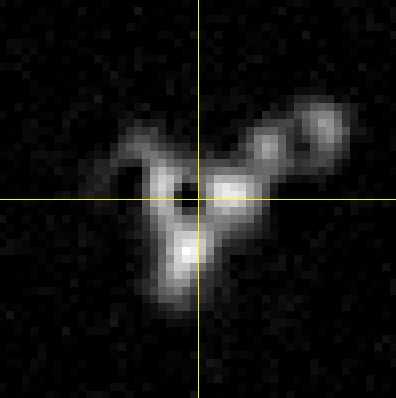} &
\includegraphics[width=50pt]{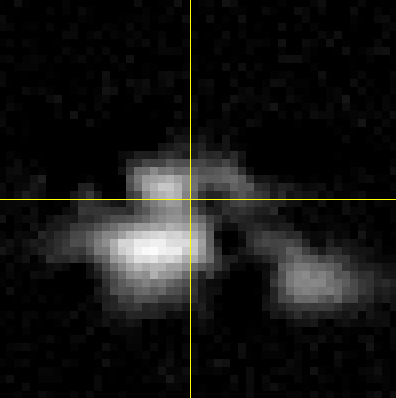} &
\includegraphics[width=50pt]{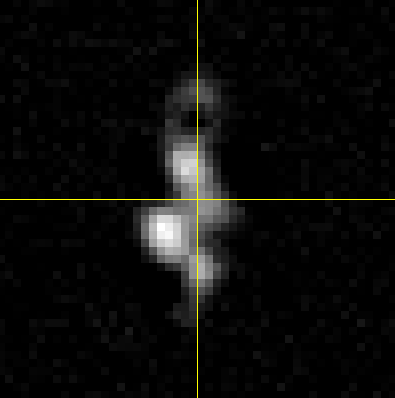} &
\includegraphics[width=50pt]{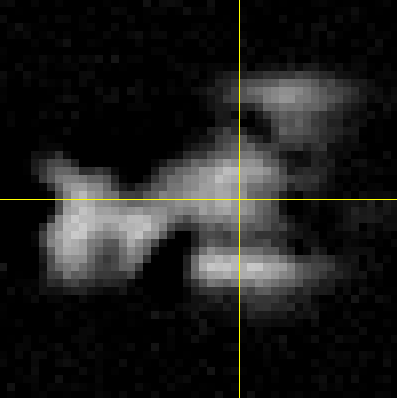}
\\[-2pt]
&\includegraphics[width=50pt]{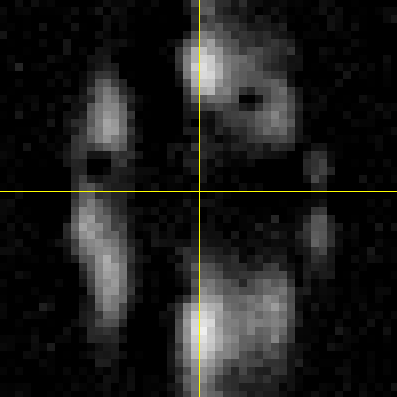} &
\includegraphics[width=50pt]{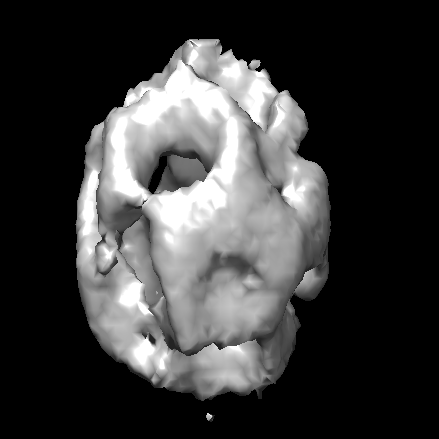} &
\includegraphics[width=50pt]{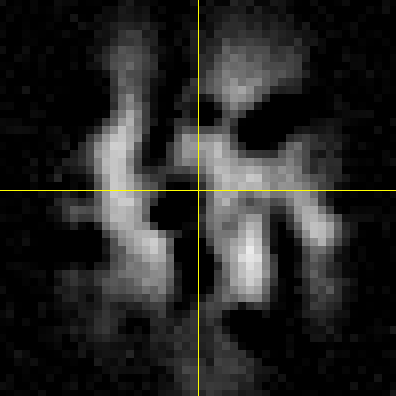} &
\includegraphics[width=50pt]{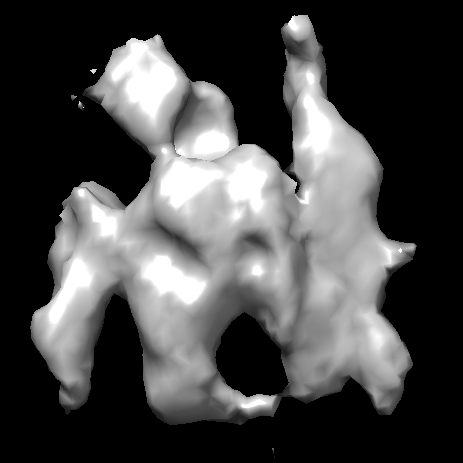}&
\includegraphics[width=50pt]{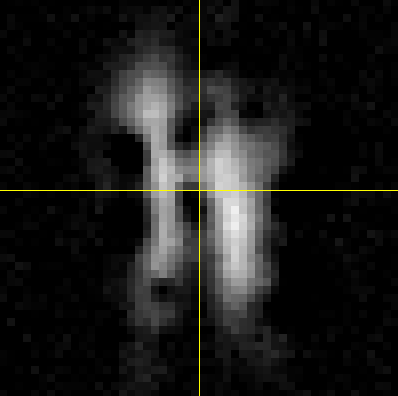} &
\includegraphics[width=50pt]{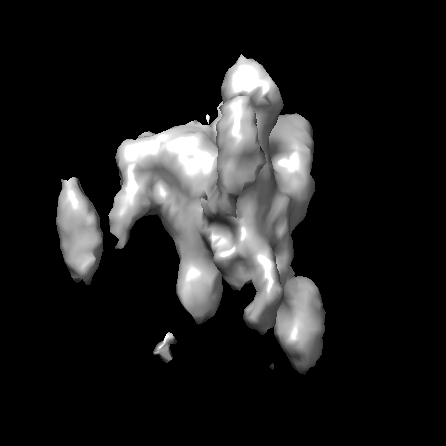}&
\includegraphics[width=50pt]{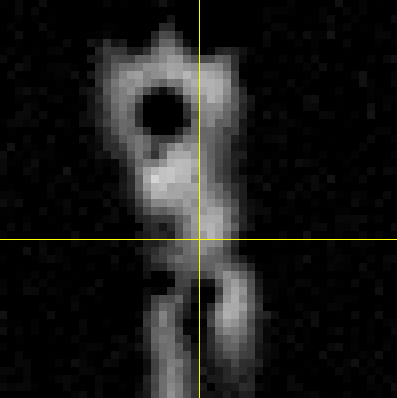} &
\includegraphics[width=50pt]{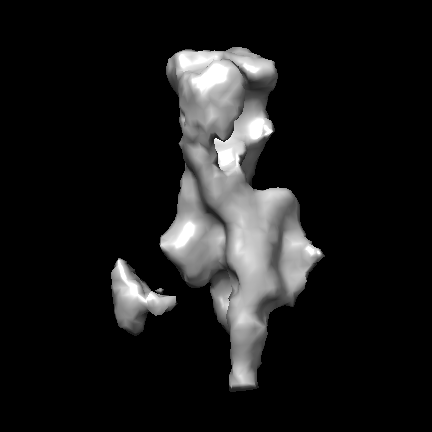}
\end{tabular}
\caption{Examples of simulated data. Top rows: ground truth volume; bottom rows: example of a view simulated from the ground truth. For each volume, we show 3 orthogonal views and a 3D isosurface visualization.}
\label{fig:visu_views}
\end{figure}

We use particle models obtained by reconstruction in cryo-EM and available in the databank EMDataRessource\footnote{https://www.emdataresource.org/}. We select the particles AMPA receptor \cite{Zhao18}, NLR resistosome  \cite{Wang19}, Clathrin  \cite{Morris19}, HIV-vaccine \cite{cirelli19}. We consider these models as ground truth and simulate the convolutional acquisition process \eqref{mod_obs_eq} to obtain a set of views, with random orientations. We used a Gaussian PSF $h$ with covariance $\Sigma_h$ defined by: 
\begin{equation}
  \Sigma_{h} = \begin{pmatrix} \sigma_{xy}^2 & 0 & 0 \\ 0 & \sigma_{xy}^2 & 0 \\ 0 & 0 & \sigma_{z}^2
\end{pmatrix}, 
\end{equation}
where $\sigma_{z}$ and $\sigma_{xy}$ are the standard deviations along and perpendicular to the microscope axis, respectively. To mimic a realistic anisotropic blur, we choose $\sigma_{z} = 5$ and $\sigma_{xy}=1.5$ (unless stated otherwise). Note that we choose this particular PSF model because it corresponds to a realistic PSF, but our method can be applied to any PSF. The ground truth volumes are resized to $50 \times 50 \times 50$ pixels before generating the views. We apply a Gaussian additive noise with standard deviation $0.2$ (images are scaled from $0$ to $1$). Intensity variations due to non uniform DOL are simulated by subtracting small Gaussian spots (3D uniform Gaussian functions) at random locations in the image. The standard deviation of the Gaussian spots is uniformly sampled between 2 to 5 percent of the image size, the intensity of the spots is uniformly sampled between 0 and 1 and the number of spots to be removed is set to 120.
On Figure \ref{fig:visu_views}, we show the ground truth volumes and examples of simulated views, aligned with the ground truth. To represent them, we use two complementary types of visualizations: three orthogonal views (planes \XY, YZ and XZ), and an isosurface 3D visualization (with Chimera \footnote{https://www.cgl.ucsf.edu/chimera/}). The anisotropy of resolution can be visualized on the intensity maps XZ and YZ.    

\subsection{Metrics}

\label{part:metrics}
 
We use three evaluation metrics to evaluate the performance of our method on simulated data. To quantify the visual quality of the reconstruction, we compute the Structural Similarity Index (SSIM) \cite{ssim} between the ground truth image and the reconstructed volume.

To measure the resolution of the reconstruction, we follow the standard practice in electron microscopy and compute the Fourier Shell Correlation (FSC) \cite{van2005fourier}. The FSC between the reconstruction $f^*$ and the ground truth $f_{gt}$ is a function $FSC: \R \rightarrow \R$ defined by: 
\begin{equation}
FSC(r) = \frac{\sum_{x \in \mathcal{S}_r} \mathcal{F}(f ^ *)(x)\overline{\mathcal{F}(f_{gt})}(x)}{\sqrt{\sum_{x \in \mathcal{S}_r} \mathcal{F}(f ^ *) ^ 2(x) \sum_{x \in \mathcal{S}_r} \mathcal{F}(f_{gt}) ^ 2 (x)}}
\end{equation}
where $\mathcal{S}_r$ denotes the sphere of radius $r$. Thus, the FSC measures the normalized cross-correlation between spherical shells of $f^ *$ and $f_{gt}$ at different frequency magnitudes in the Fourier space. The resolution of $f^ *$ can be derived from the FSC by determining the frequency limit after which the correlation becomes lower than a cutoff value (we consider a standard cutoff of 0.143). In what follows, we will use the name FSC to refer to to the resolution derived from the FSC curve.

In our case, since the resolution of the input data is anisotropic, it is important to assess the ability of our method to make it isotropic. To quantify the degree of isotropy of the resolution, we use a variant of the FSC named conical FSC (cFSC) \cite{DIEBOLDER2015215}. The principle of the cFSC is to compute a FSC curve locally for each direction. Thus, we can derive a resolution value for several directions and plot them on a map. Each point on the map is a direction represented by two angles $\phi_1 \in [0,2\pi]$ and $\phi_2 \in [0,\pi]$ (see Figure \ref{spherical coordinates}). On Figure~\ref{fig:conical_fsc_map_view} we show the cFSC map between the ground truth object AMPA receptor and a generated view (aligned with the ground truth). This figure highlights the anisotropy of resolution: the resolution in the Z axis ($\phi_2 = 0$ and $\phi_2 = 180$) is much lower than in other directions. The resolution values are expressed as the inverse of the pixel size.     

Since the orientation of the reconstruction does not match the orientation of the ground truth object, the reconstructed volume is registered on the ground truth before computing the FSC, cFSC or the SSIM. The registration is performed with a first step of exhaustive search method maximizing the mutual information. The discretization step is set to 10 degrees. To refine the registration we then use a gradient descent. 

\subsection{Reconstruction results on simulated data}

\begin{figure}
    \centering
    \begin{subfigure}[b]{0.48\textwidth}
        \includegraphics[width=\textwidth]{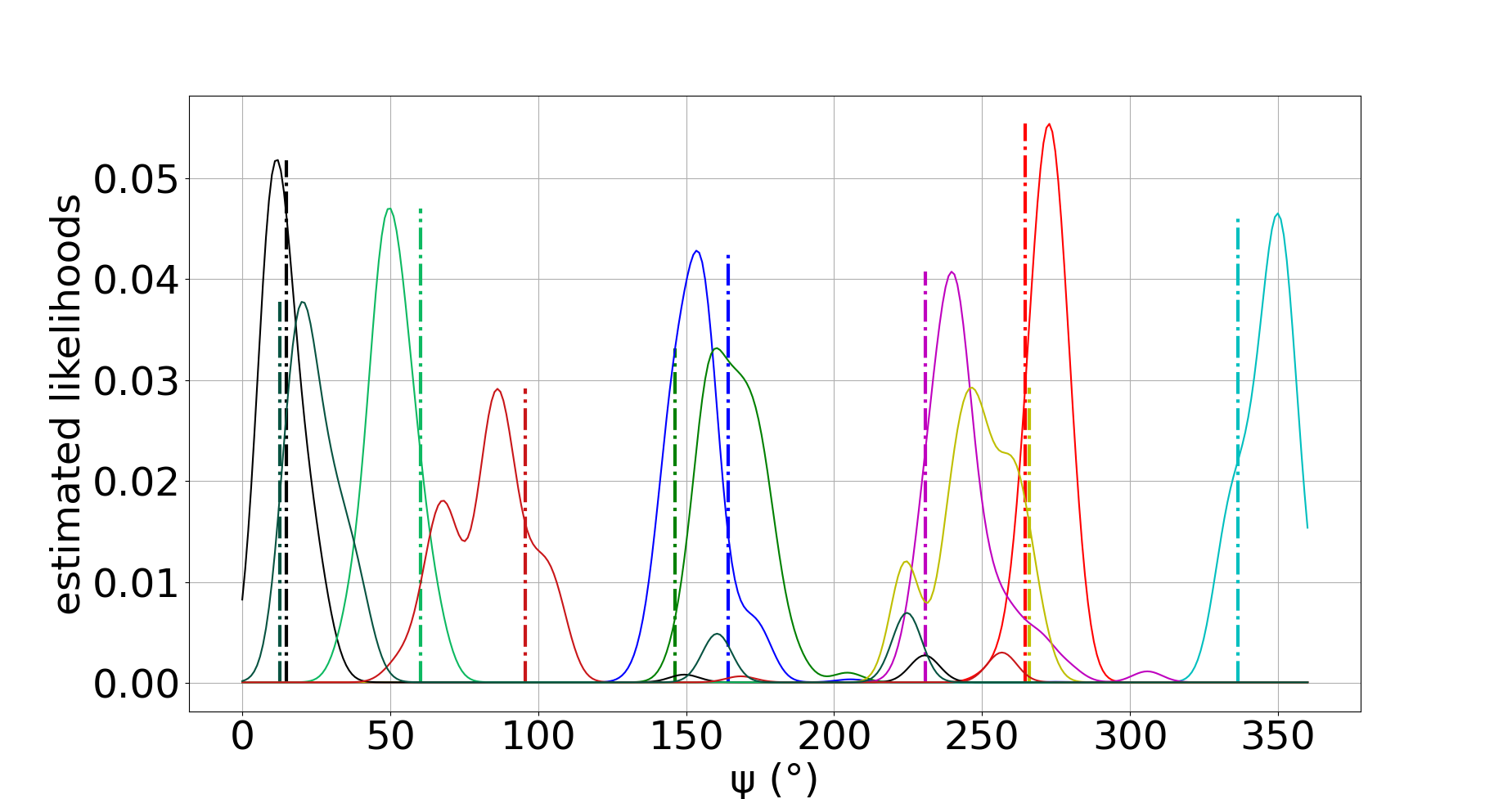}
        \caption{First 10 views.}
        \label{fig:lik_psi0}
    \end{subfigure}
    \begin{subfigure}[b]{0.48\textwidth}
        \includegraphics[width=\textwidth]{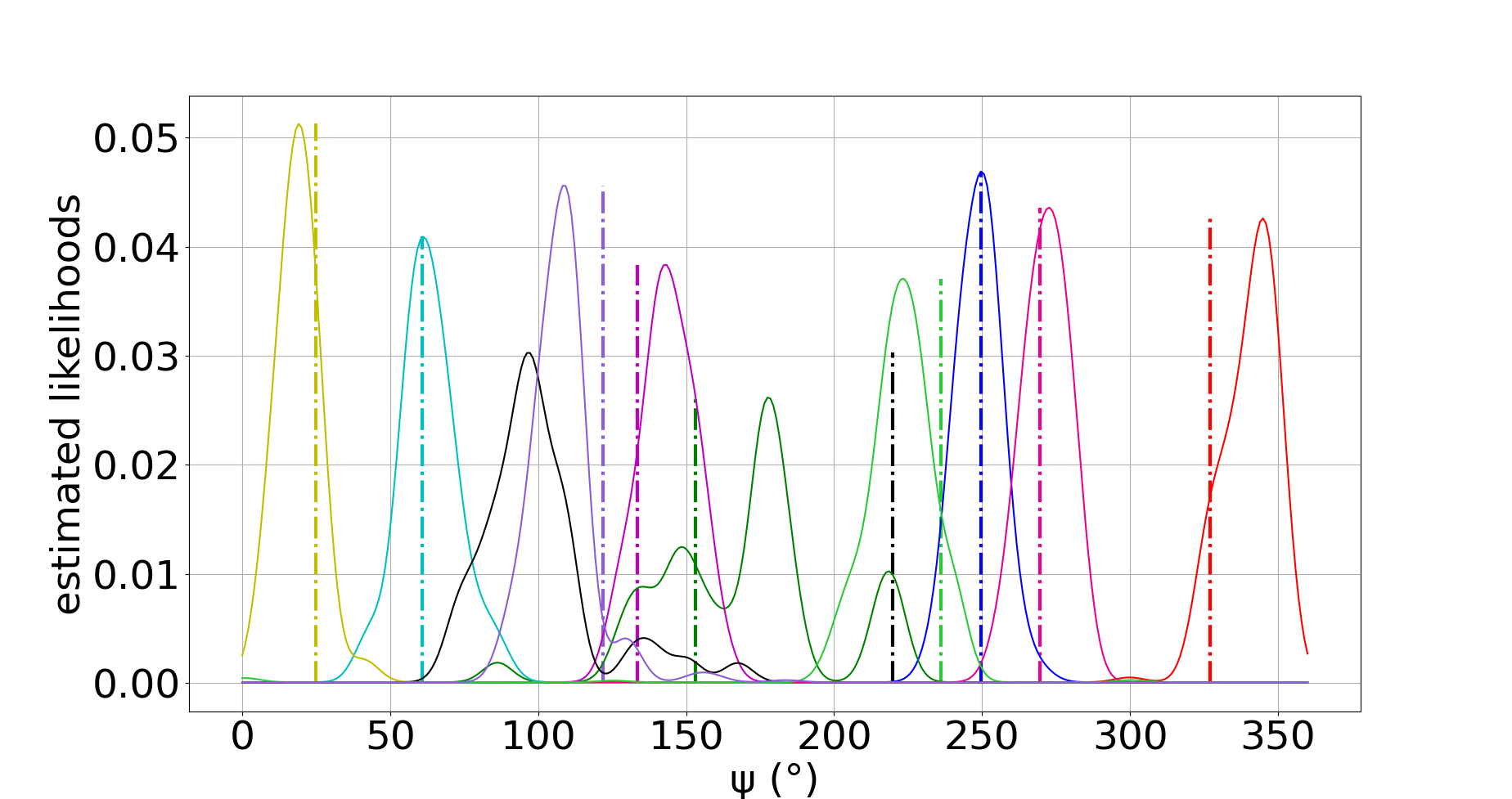}
        \caption{Last 10 views.}
        \label{fig:lik_psi}
    \end{subfigure}\\
    \begin{subfigure}[b]{0.6\textwidth}
        \includegraphics[width=\textwidth]{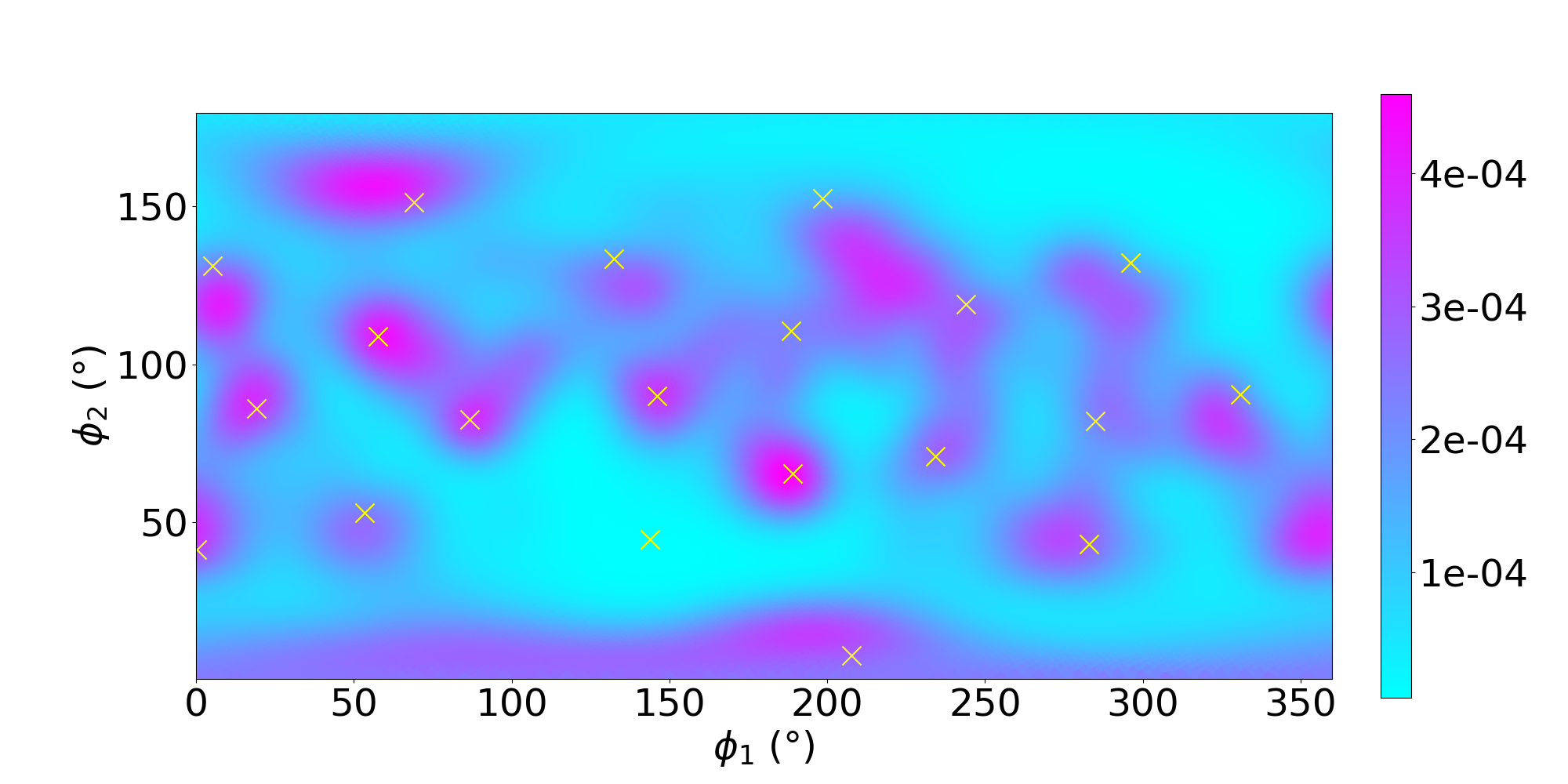}
        \caption{}
        \label{fig:lik_d}
    \end{subfigure}

    \caption{Approximated likelihoods $\alk^ {l, \psi}$ and $\alk^ {l, d}$ at the last iteration of the reconstruction on data simulated from the AMPA receptors (see Figure \ref{fig:visu_views}). (a) and (b): Likelihoods $\alk^ {l, \psi}$, where each curve represents a view and the vertical bars represent the true values (the results are split in two plots for a better visualization). (c): Sum of the likelihoods $\alk^ {l, d}$ of all the views. The yellow crosses are the true values of $d$.}
\end{figure} 

\subsubsection{Experimental setting}
\label{part:exp_settings}

In all our experiments, the volume $\hat f$ is initialized with random values uniformly distributed between 0 and 1.  In the default experimental settings, we used 20 views for reconstruction.
The standard approach for reference-free particle averaging in fluorescence imaging is the application of cryo-EM reconstruction methods on 2D data. To apply this idea, we create artificial 2D projections from our 3D input volumes, and perform reconstruction with a state-of-the-art cryo-EM reconstruction method based on the RANSAC algorithm \cite{Javier2014} (integrated in the Scipion software \cite{scipion}). We name this method cryo-RANSAC in what follows. 
We also compare with the results of the SPR method dedicated to 3D convolutional models described in \cite{Fortun18}. We name it MP3DR (Multiple Particles 3D Reconstruction) for convenience. This method requires the selection of a small number of particles to maintain a tractable computational cost. We selected 5 representative particles manually for each dataset.
Finally, in order to evaluate the impact of orientation estimation errors, we  show  reconstructions results obtained with known poses, using the gradient-based reconstruction method presented in \ref{part:opt_volume}. 

\subsubsection{Estimation of the orientations}
\label{part:imp_distr}

In order to evaluate the accuracy of the orientation estimation, we show the approximated marginal likelihoods $\alk^ {l,d}$ and $\alk ^ {l, \psi}$ obtained in the last iteration of the reconstruction. We recall that these distributions are used to sample several axes $d$ and angles $\psi$, from which the optimal angle axis-angle representation is estimated. The likelihoods $\alk ^ {l,\psi}$ are shown on Figure \ref{fig:lik_psi0} and \ref{fig:lik_psi}, where each colored curve represents the likelihood of one view. The vertical bars represent the ground truth values of $\psi$ for each view. The sum of the likelihoods $\alk ^ {l, d}$ of all the views is shown on Figure \ref{fig:lik_d} in a 2D map where each axis represents a dimension of $d=(\phi_1,\phi_2)$. The ground truth values of $d$ are represented with yellow crosses. 

For both $\psi$ and $d$, we observe that the likelihoods are approximately centered around the true values, which shows that the orientations are well estimated. Note that we aim at a coarse {\it ab initio} reconstruction that is destined to serve as an initialization for a refinement algorithm \cite{Fortun18}. Only one estimated likelihood is not located close to the true value (the black curve on Figure \ref{fig:lik_psi}).

\subsubsection{Quantitative reconstruction results}
\label{sec:quantitative_results}

\begin{table*}[t]
\begin{center}
\begin{tabular}{|c|cc|cc|cc|cc|}
    \hline
    &\multicolumn{2}{c|}{Known poses}&\multicolumn{2}{c|}{Our method}&\multicolumn{2}{c|}{cryo-RANSAC}&\multicolumn{2}{c|}{{MP3DR}}\\
    & SSIM & FSC & SSIM & FSC  & SSIM & FSC & SSIM & FSC \\
	\hline
    AMPA receptors&0.91&0.42&0.82&0.27&0.54&0.05&{0.72}&{0.14}\\
    HIV Vaccine&0.93&0.44&0.83&0.28&0.53&0.06&0.72&0.13\\
    clathrine&0.92&0.34&0.82&0.29&0.55&0.11&0.67&0.14\\
    NLR resistosome & 0.93&0.35&0.88&0.28&0.53&0.10&0.81&0.17\\
    \hline
\end{tabular}
\end{center}
\caption{Quantitative comparison of the results of our method, cryo-RANSAC, MP3DR and the reconstruction with known poses.}
\label{table_SSIM}
\end{table*}

\begin{figure}
     \centering
     \begin{subfigure}[b]{0.33\textwidth}
         \centering
         \includegraphics[width=\textwidth]{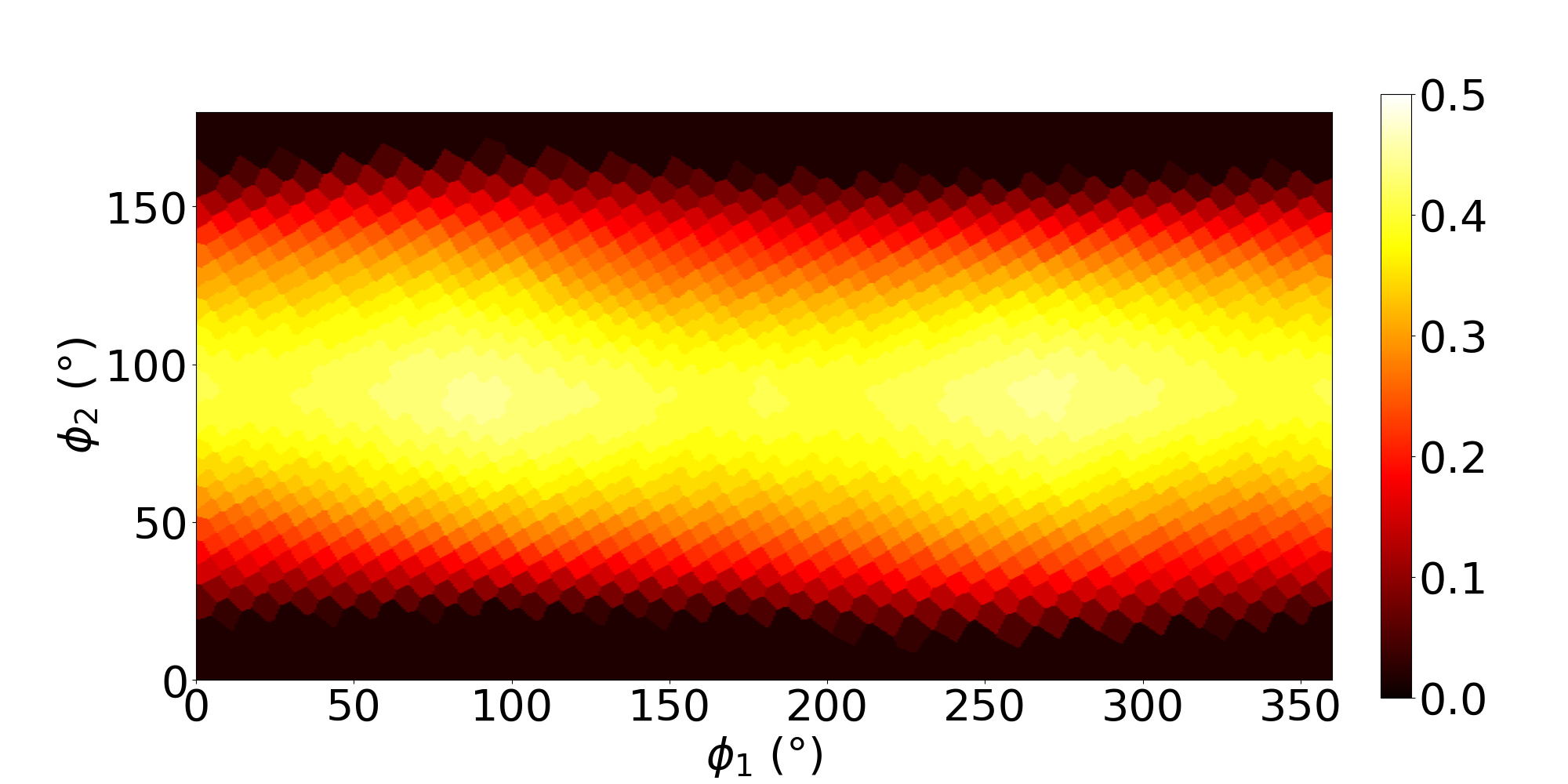}
         \caption{Input view}
     \label{fig:conical_fsc_map_view}   
     \end{subfigure}
     \begin{subfigure}[b]{0.33\textwidth}
         \centering
         \includegraphics[width=\textwidth]{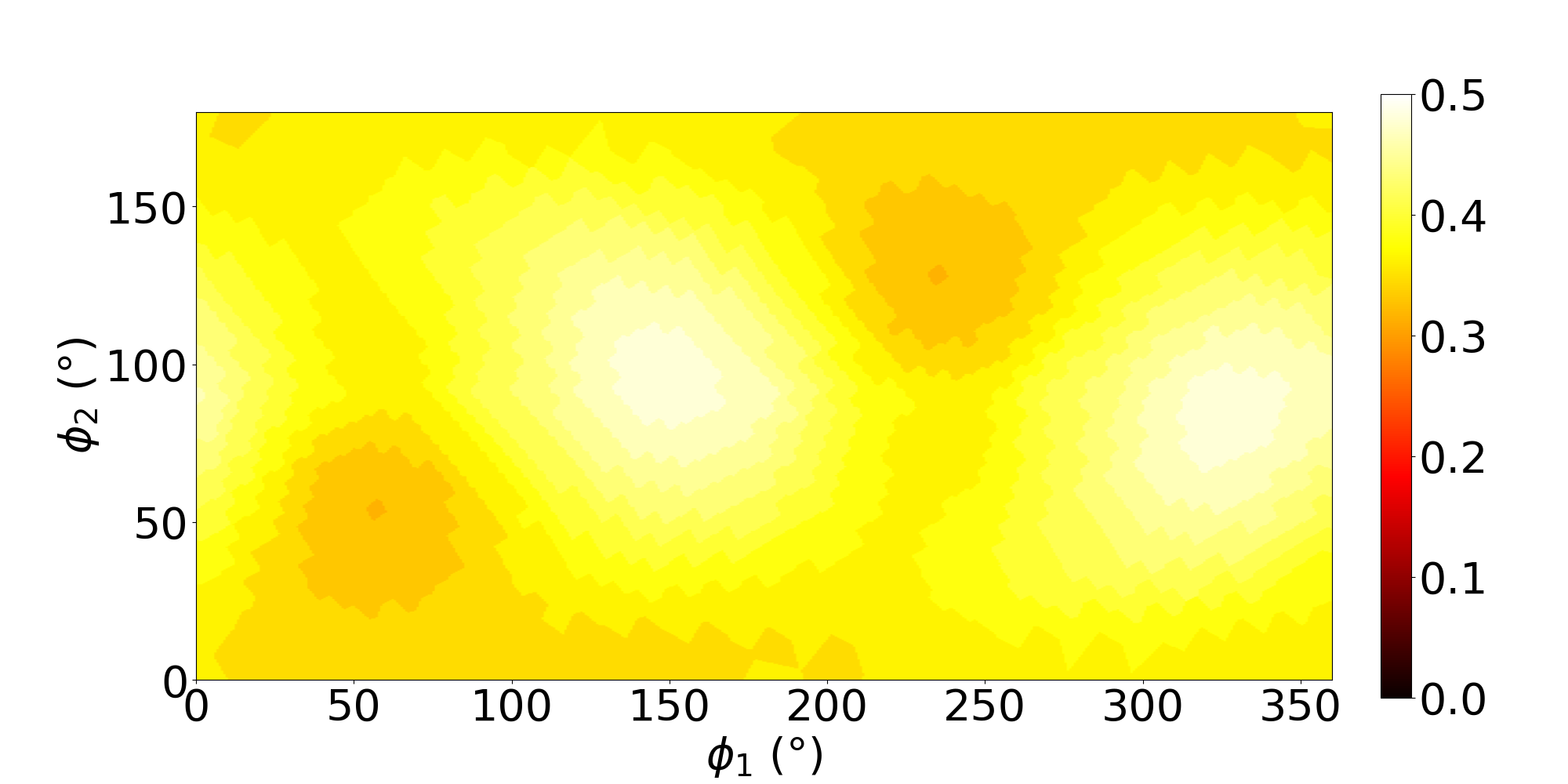}
         \caption{Our method}
	\label{fig:conical_fsc_map_ours}     
     \end{subfigure}
     \begin{subfigure}[b]{0.33\textwidth}
         \centering
         \includegraphics[width=\textwidth]{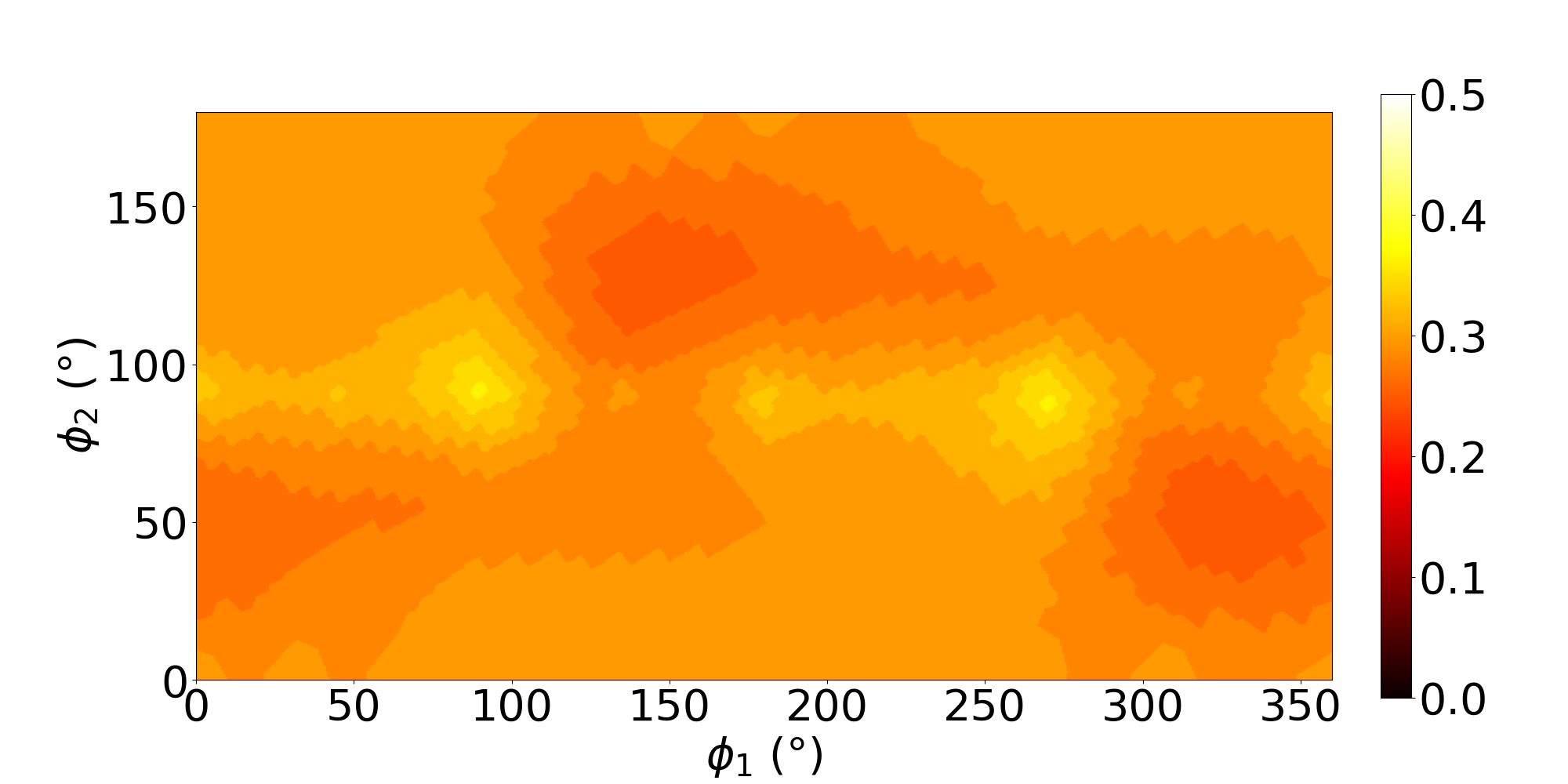}
         \caption{cryo-RANSAC}
	\label{fig:conical_fsc_map_ransac}     
     \end{subfigure}

\caption{cFSC maps between the ground truth object (here the AMPA receptor) and a view \ref{fig:conical_fsc_map_view}, our reconstruction \ref{fig:conical_fsc_map_ours} and cryo-RANSAC reconstruction \ref{fig:conical_fsc_map_ransac}. Each point on the map is a direction, represented by two angles $\phi_1 \in [0, 360]$ and $\phi_2 \in [0,180]$ (see Figure \ref{spherical coordinates}). }
\label{fig:cFSC_map}    
\end{figure}

\begin{figure*}
\centering
\begin{tabular}{m{10pt}@{\hspace{5pt}}m{74pt}@{\hspace{10pt}}m{74pt}@{\hspace{10pt}}m{74pt}@{\hspace{10pt}}m{74pt}@{\hspace{10pt}}m{74pt}}
&\centering {\bf Ground truth}&\centering {\bf Known poses}&\centering {\bf Our method}&\centering {\bf cryo-RANSAC}& ~~~~~{\bf {MP3DR}}\\

\multirow{2}{*}{\centering \rotatebox{90}{Clathrin~~~~~~}}
&\includegraphics[width=74pt]{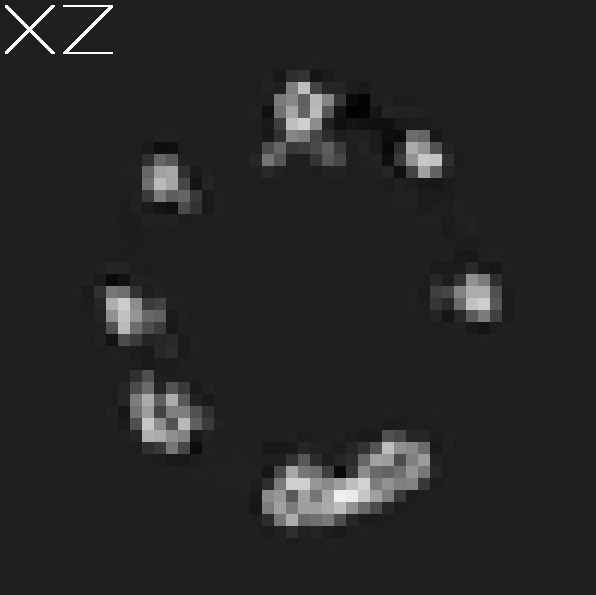}
&\includegraphics[width=74pt]{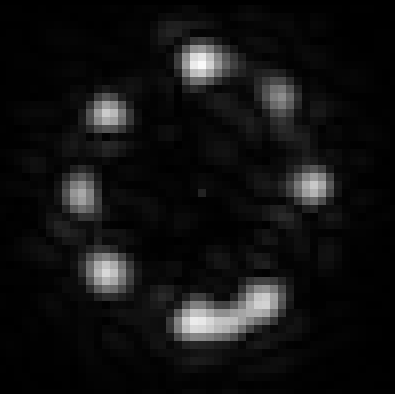}
&\includegraphics[width=74pt]{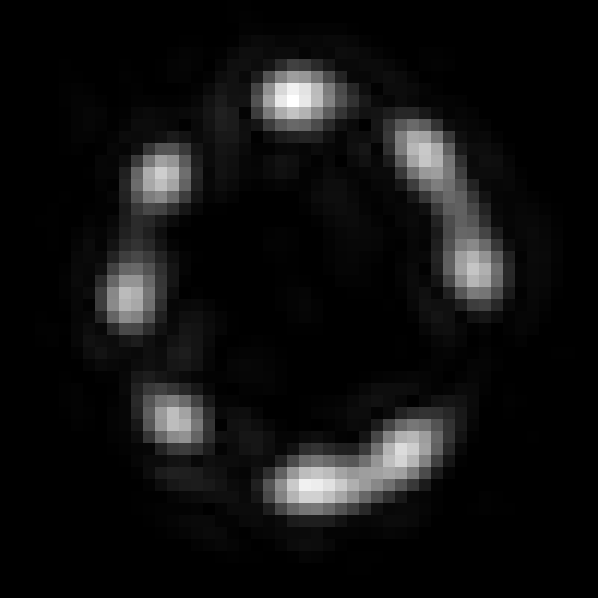}
&\includegraphics[width=74pt]{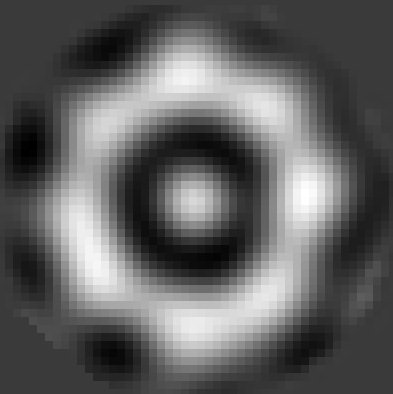}
&\includegraphics[width=74pt]{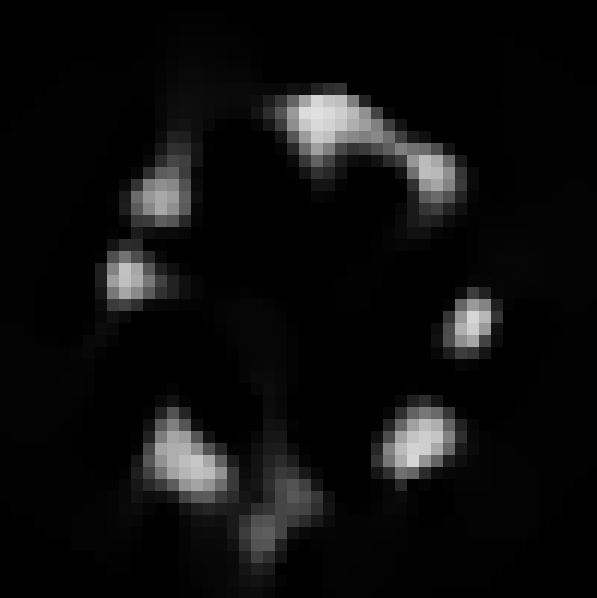}\\[-2pt]

&\includegraphics[width=74pt]{illustrations/illustr_views/clathrine/gt/chimera.png}
&\includegraphics[width=74pt]{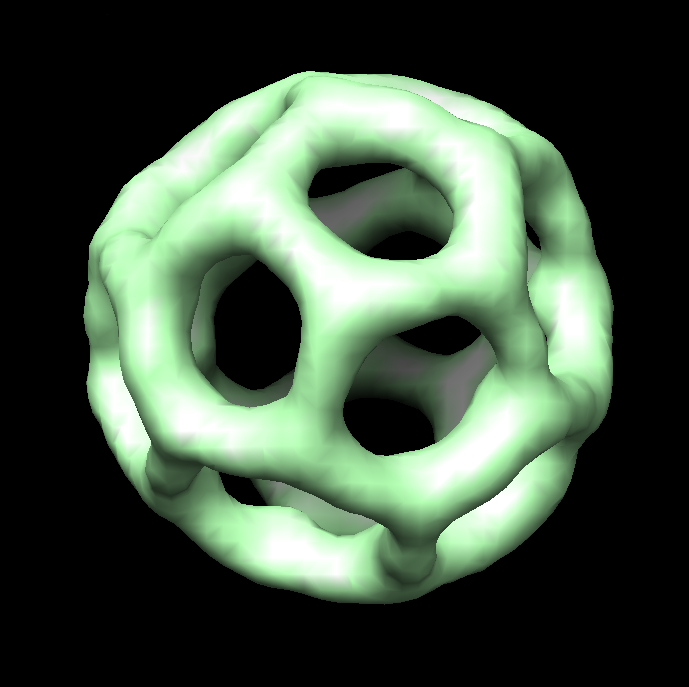}
&\includegraphics[width=74pt]{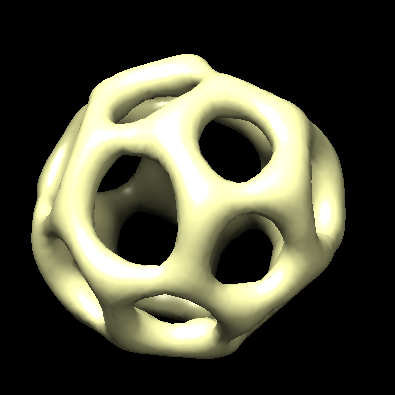}
&\includegraphics[width=74pt]{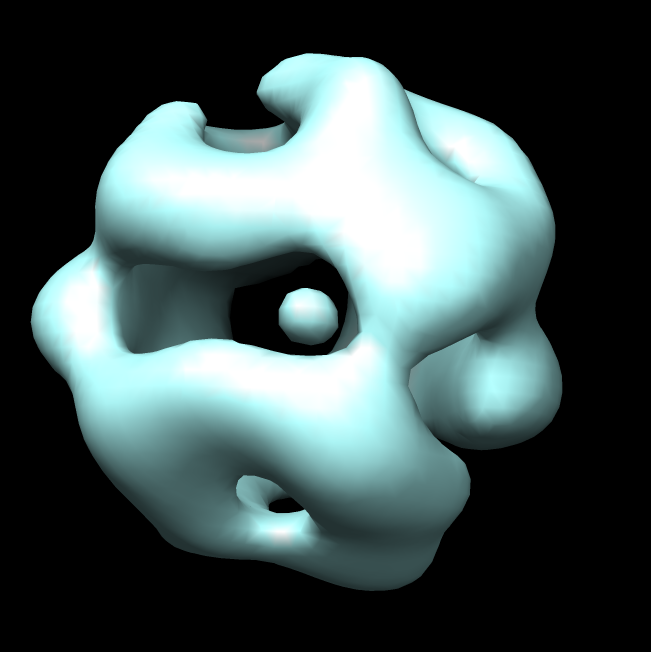}
&\includegraphics[width=74pt]{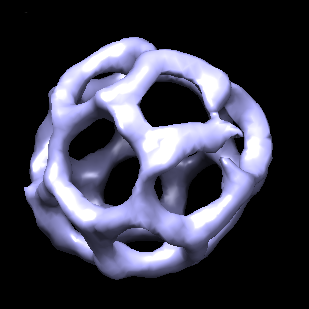}\\[-3pt]

\midrule[1.5pt] \\[-12pt]
\multirow{2}{*}{\centering \rotatebox{90}{NLR resistosome~~~}}
&\includegraphics[width=74pt]{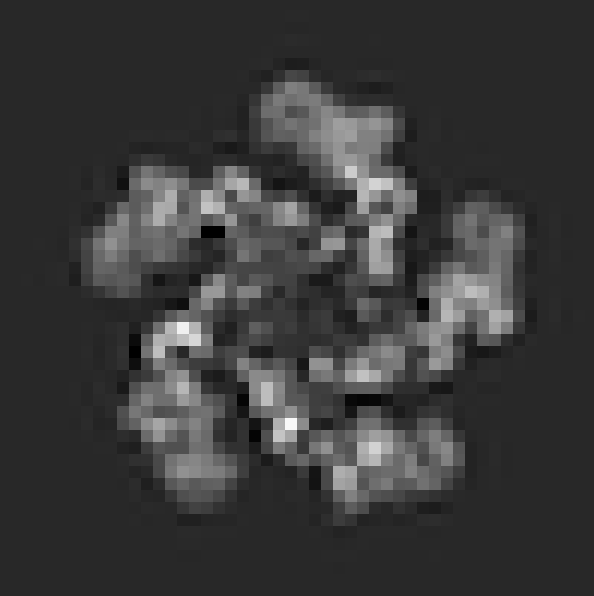}&
\includegraphics[width=74pt]{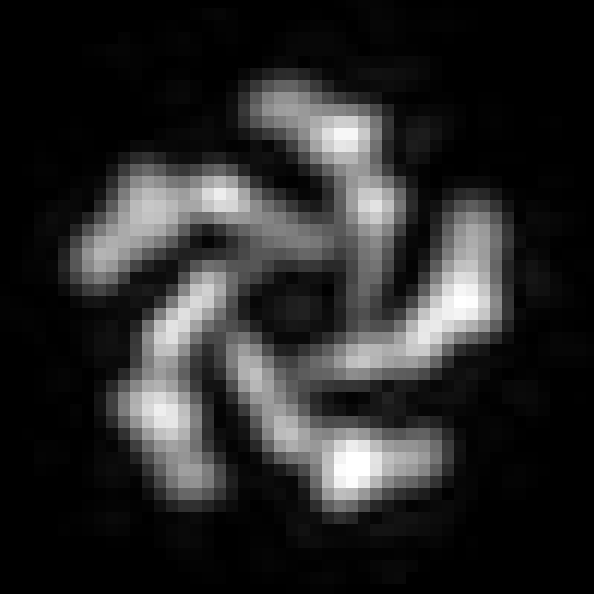}
&\includegraphics[width=74pt]{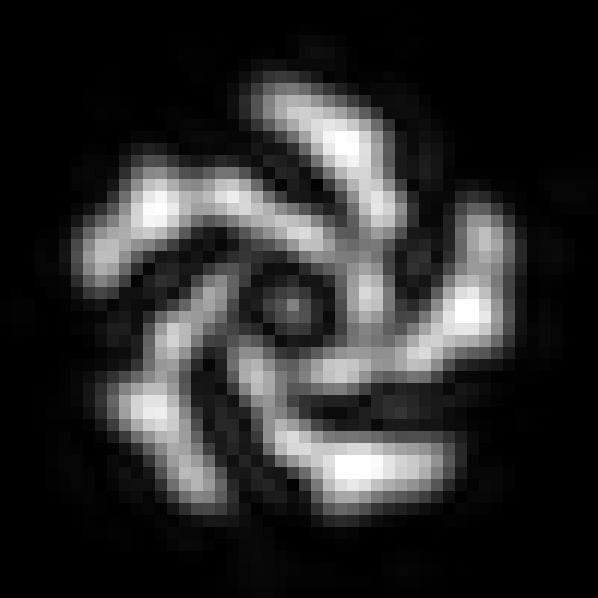}
&\includegraphics[width=74pt]{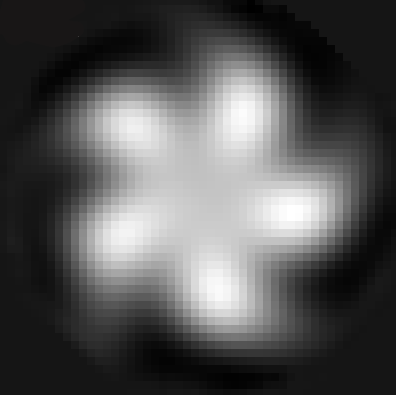}
&\includegraphics[width=74pt]{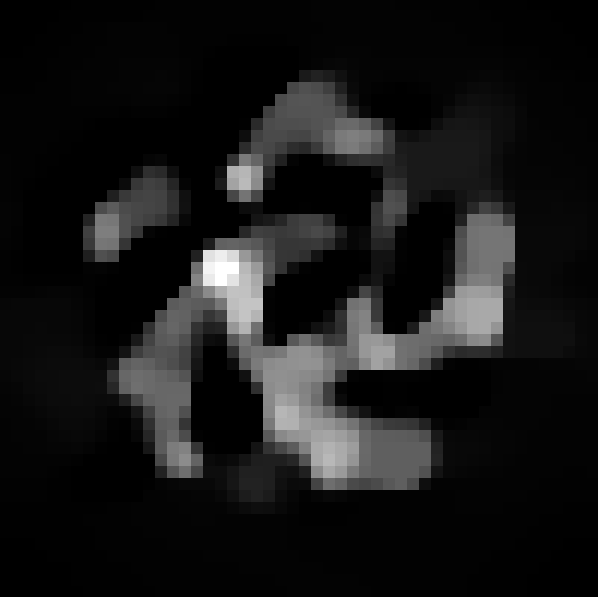} \\[-2pt]

&\includegraphics[width=74pt]{illustrations/illustr_views/emd_0680/gt/chimera.png}
&\includegraphics[width=74pt]{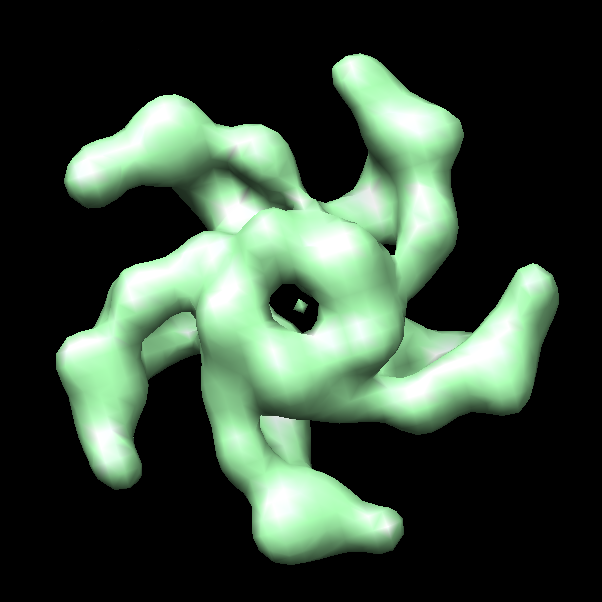}
&\includegraphics[width=74pt]{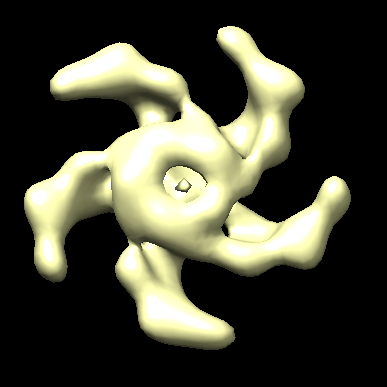}
&\includegraphics[width=74pt]{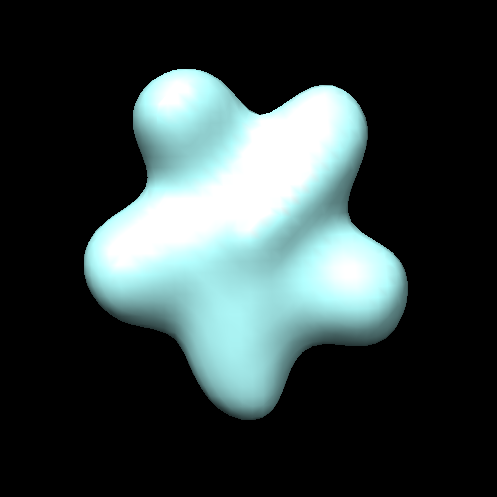}
&\includegraphics[width=74pt]{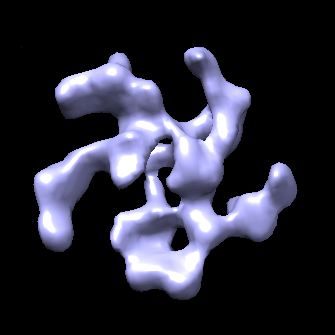}\\[-3pt]

\midrule[1.5pt] \\[-12pt]
\multirow{2}{*}{\centering \rotatebox{90}{HIV-Vaccine~~~~~}}
&\includegraphics[width=74pt]{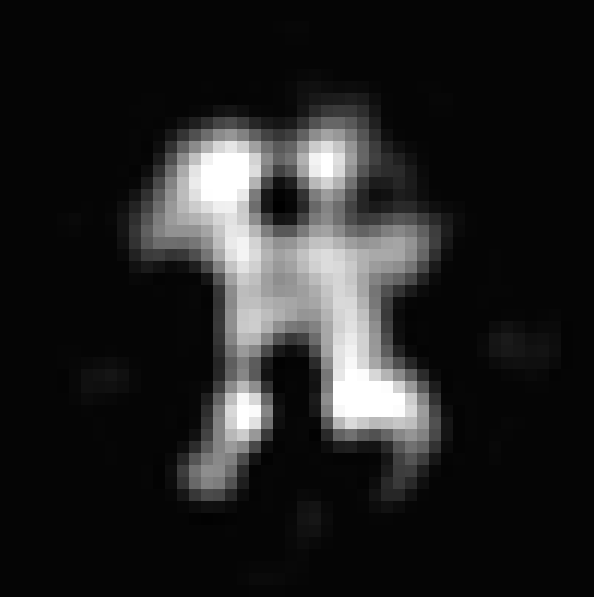}&
\includegraphics[width=74pt]{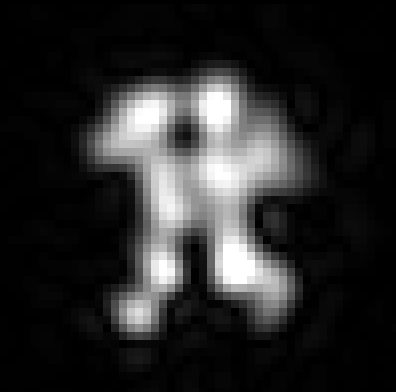}
&\includegraphics[width=74pt]{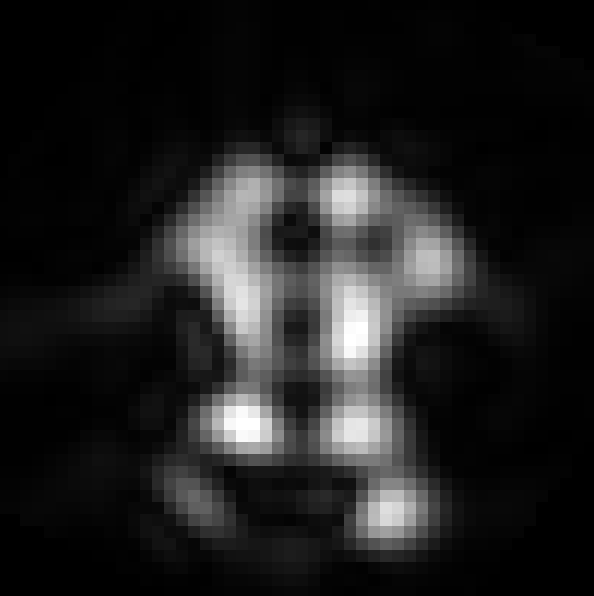}
&\includegraphics[width=74pt]{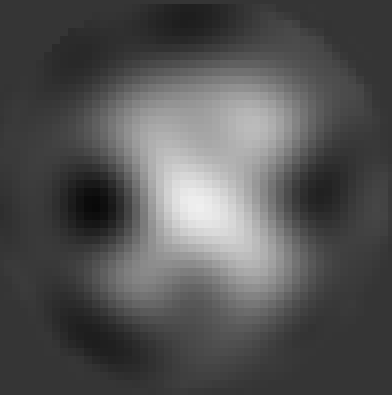}
&\includegraphics[width=74pt]{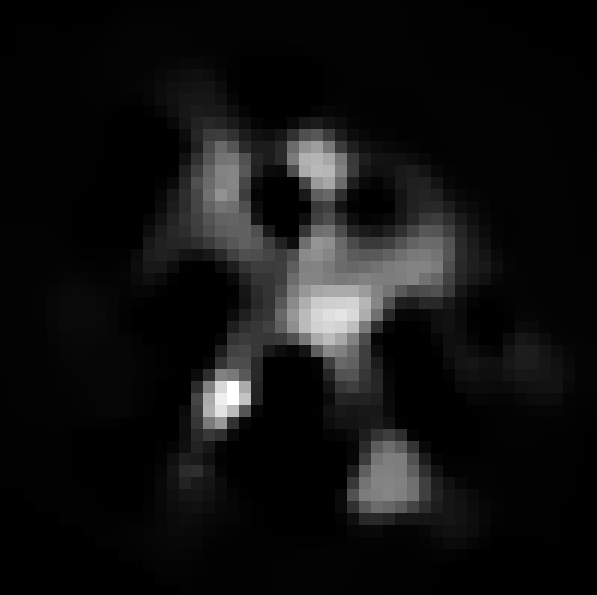}\\[-2pt]

&\includegraphics[width=74pt]{illustrations/illustr_views/HIV_vaccine/gt/chimera.png}
&\includegraphics[width=74pt]{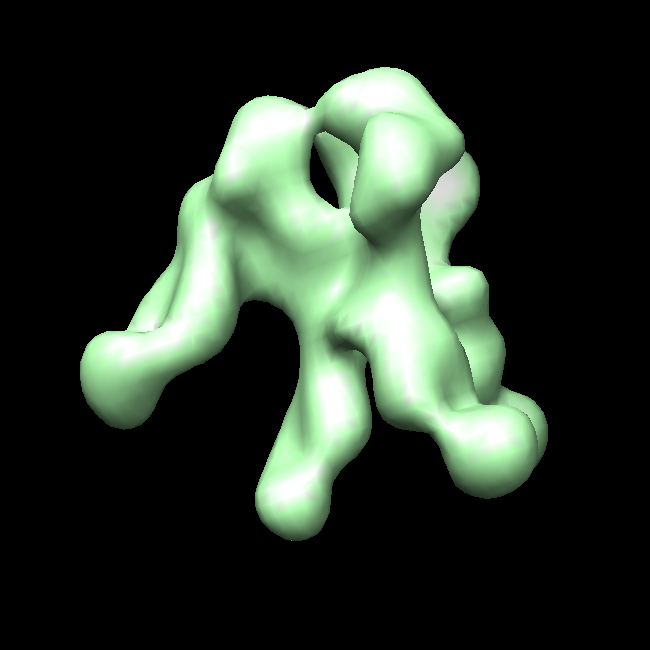}
&\includegraphics[width=74pt]{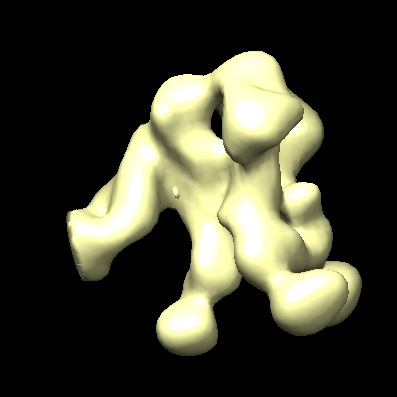}
&\includegraphics[width=74pt]{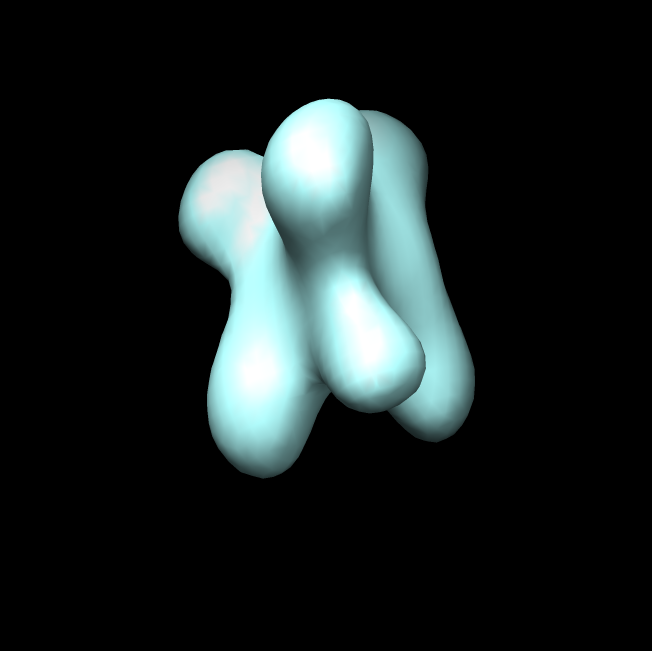}
&\includegraphics[width=74pt]{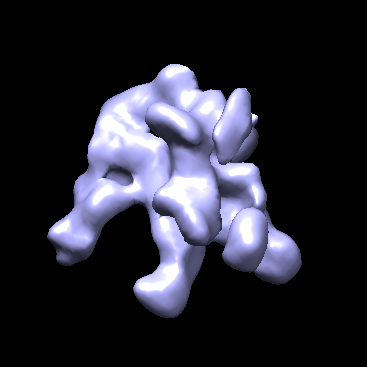}\\[-3pt] 

\midrule[1.5pt] \\[-12pt]
\multirow{2}{*}{\centering \rotatebox{90}{AMPA receptors~~~~}}
&\includegraphics[width=74pt]{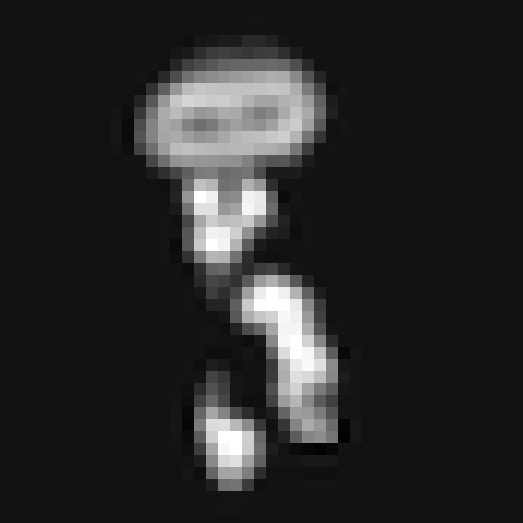}&
\includegraphics[width=74pt]{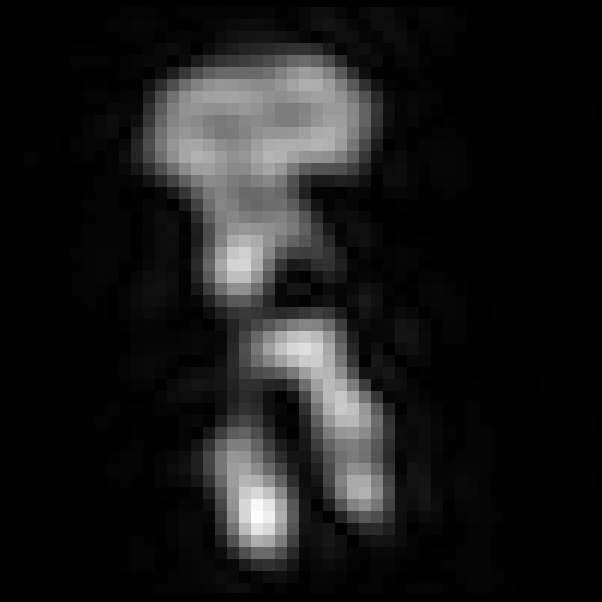}
&\includegraphics[width=74pt]{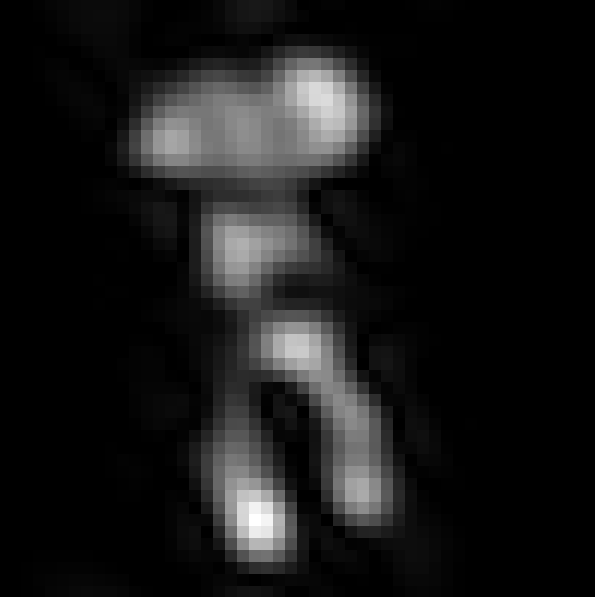}
&\includegraphics[width=74pt]{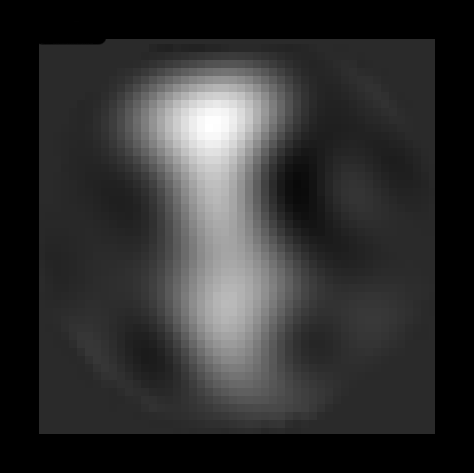}
&\includegraphics[width=74pt]{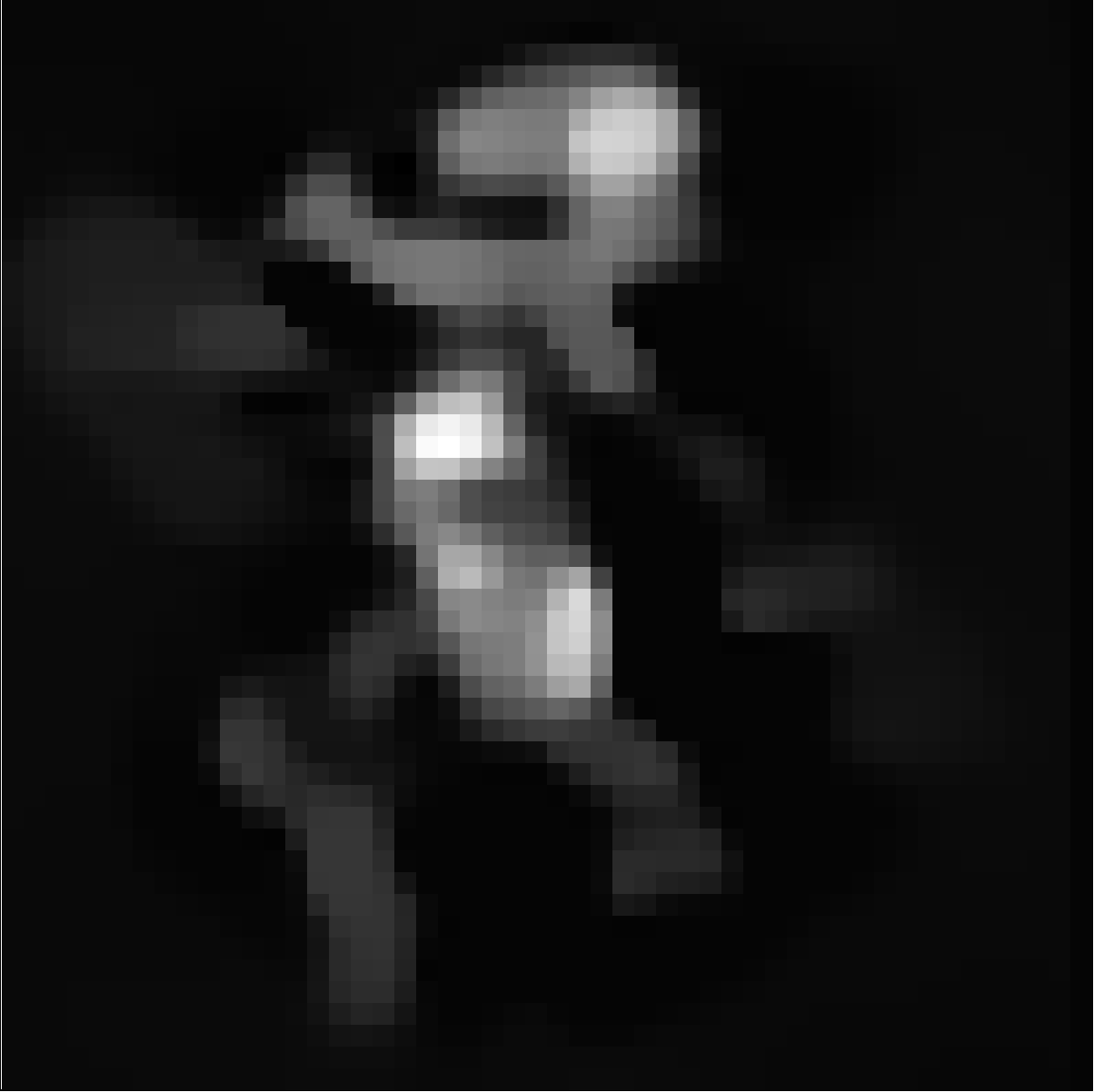}\\[-2pt]

&\includegraphics[width=74pt]{illustrations/illustr_views/recepteurs_AMPA/gt/chimera.png}
&\includegraphics[width=74pt]{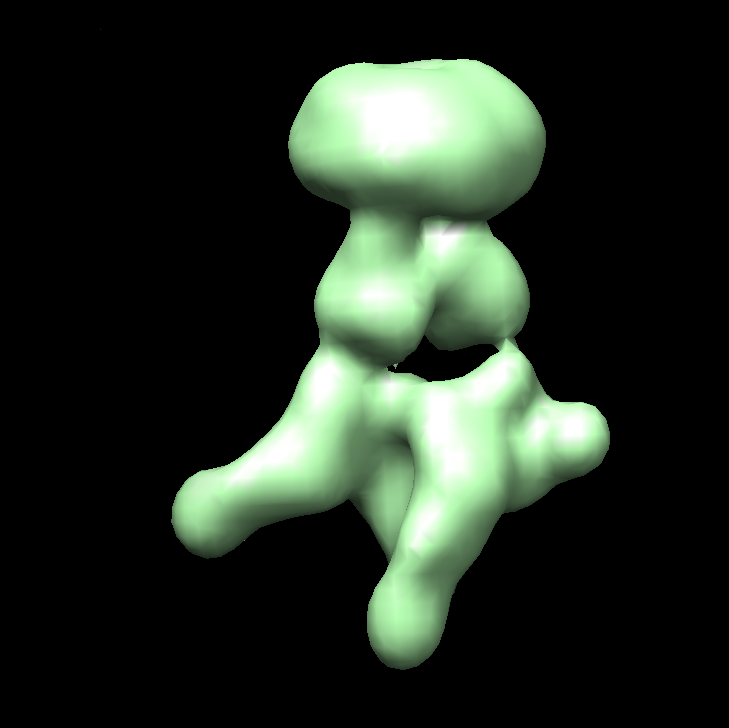}
&\includegraphics[width=74pt]{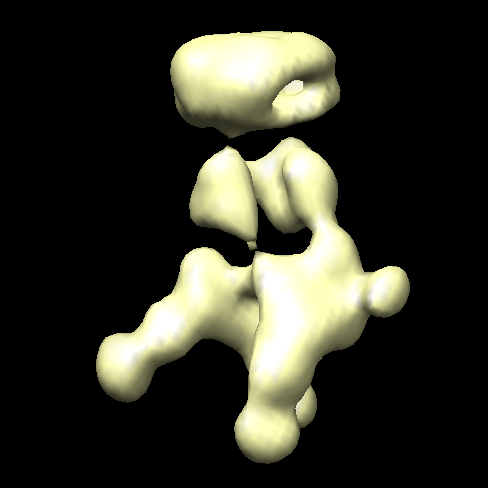}
&\includegraphics[width=74pt]{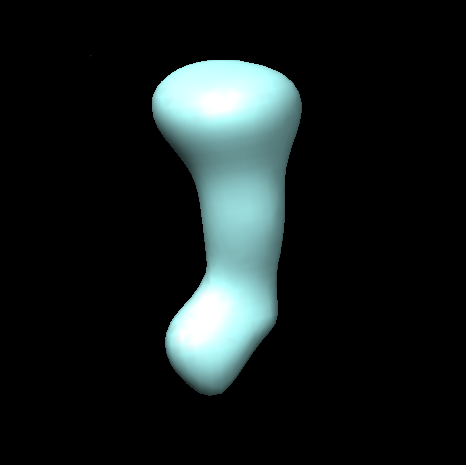}
&\includegraphics[width=74pt]{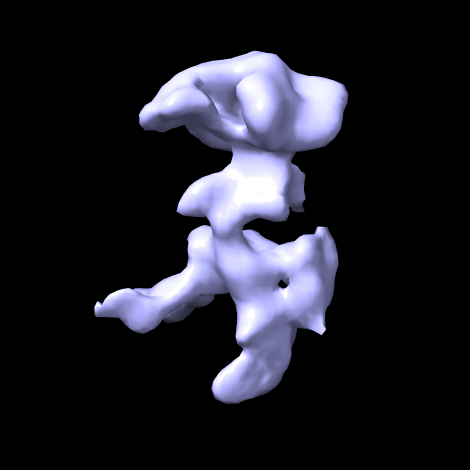}
\end{tabular}

\caption{Reconstruction results obtained on synthetic data. Each volume is represented with an orthogonal view in the XZ plane and a 3D isosurface. From left to right columns: ground truth object, reconstruction with known poses, reconstruction with our method, cryo-RANSAC reconstruction, {MP3DR} reconstruction. 20 views are used for the reconstruction.}

\label{fig:visu_recons}
\end{figure*}

Table \ref{table_SSIM} reports the SSIM and FSC values obtained on the simulated data with our algorithm, cryo-RANSAC, {MP3DR}, and the reconstruction with known poses. Since our method, as well as the generation of the views involves randomness, we perform 100 experiments for the 4 particles and report the average. 

We found that for all the particles, our algorithm provides significantly better SSIM and FSC  than cryo-RANSAC reconstructions. This result illustrates the importance to take into account an appropriate 3D imaging model, whereas working on 2D projections induces an irreversible and critical loss of information. We also consistently outperform {MP3DR} on the two considered metrics. Note that in addition to this quantitative improvement, one of the main advantages of our method compared to {MP3DR} is that we do not need to select a reduced number of views, which is subject to classification error or human bias, and we can use directly all the data for the reconstruction. As expected, our method provides slightly lower SSIM and FSC than the reconstruction with known poses. This results shows that the orientation estimation errors analyzed in Section \ref{part:imp_distr} only have a small impact on the final reconstruction.

To quantify the resolution isotropy of the reconstructions, we show on Figure \ref{fig:cFSC_map} the cFSC maps associated to a view (aligned with the ground truth image), our reconstruction and cryo-RANSAC reconstructions, for the particle AMPA receptors. We clearly see on Figure \ref{fig:conical_fsc_map_ours} that the axial resolution has been improved. Indeed, on the cFSC map associated to the view \ref{fig:conical_fsc_map_view}, the cFSC values along the Z axis ($\phi_2 = 180$ or $\phi_2 = 0$) is almost zero, while it is between $0.35$ and $0.45$ for our reconstruction. Besides, the resolution along X and Y axis ($\phi_2 = 90 \degree$) of our reconstruction is preserved since we observe similar values to the one of the view (between $0.4$ and $0.5$). Regarding the cryo-EM dedicated algorithm, the resolution along Z axis is also improved, but the resolution along X and Y axis is degraded: the cFSC values are around $0.3$, while they are around $0.4$ for the view.

\subsubsection{Visual reconstruction}

On Figure \ref{fig:visu_recons}, we present visual reconstructions obtained with the different methods. If we compare the intensity map in plane XZ between a view and our reconstruction, we clearly see that the axial resolution has been enhanced. Visually, we do not see a clear difference between our method and the reconstruction obtained with known poses. Moreover, our reconstructions are visually closer to the ground truth than the ones of cryo-RANSAC and {MP3DR}, which confirms the quantitative results of Section \ref{sec:quantitative_results}.

\subsubsection{Robustness to anisotropy} 

\begin{figure*}
\begin{tabular}{m{35pt}@{\hspace{2pt}}m{58pt}@{\hspace{2pt}}m{58pt}@{\hspace{2pt}}m{58pt}@{\hspace{2pt}}m{58pt}@{\hspace{2pt}}m{58pt}@{\hspace{2pt}}m{58pt}@{\hspace{2pt}}m{58pt}m{3pt}}

 & \centering Example of input view & \centering 5 views & \centering 10 views & \centering 20 views & \centering 30 views & \centering 40 views & \centering 60 views
 \tabularnewline
$\sigma_z = 5$ & \includegraphics[width=58pt]{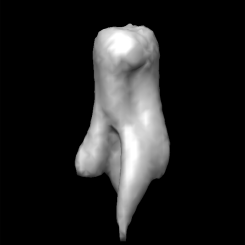} & \includegraphics[width=58pt]{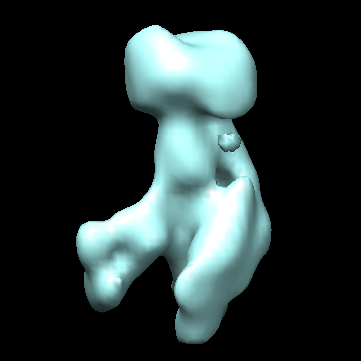} & \includegraphics[width=58pt]{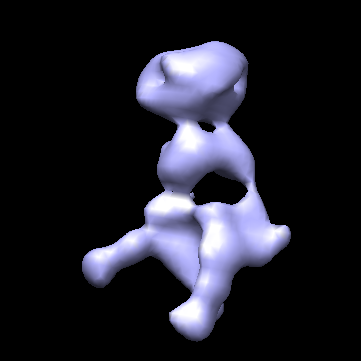} & \includegraphics[width=58pt]{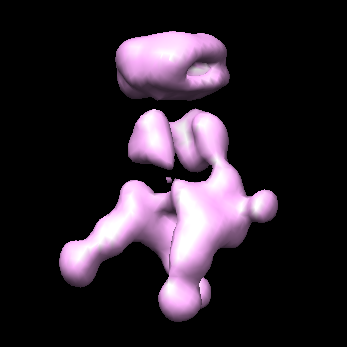} & \includegraphics[width=58pt]{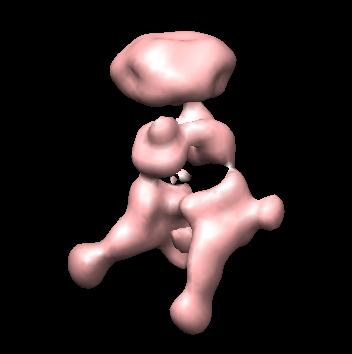} &
\includegraphics[width=58pt]{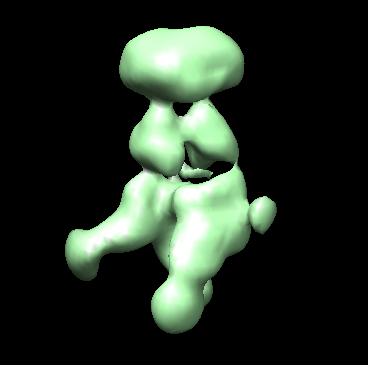} &
\includegraphics[width=58pt]{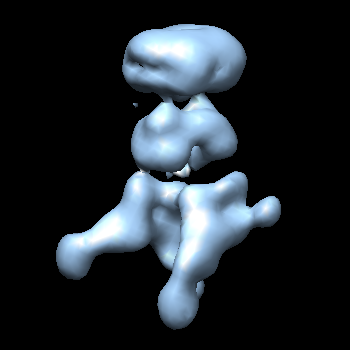} &

\tabularnewline

$\sigma_z = 10$ & \includegraphics[width=58pt]{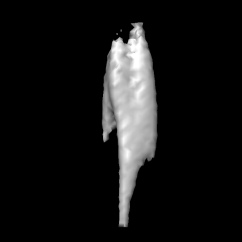} & \includegraphics[width=58pt]{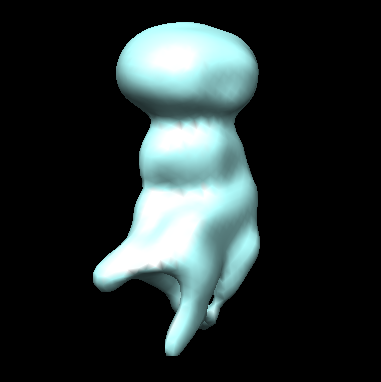} & \includegraphics[width=58pt]{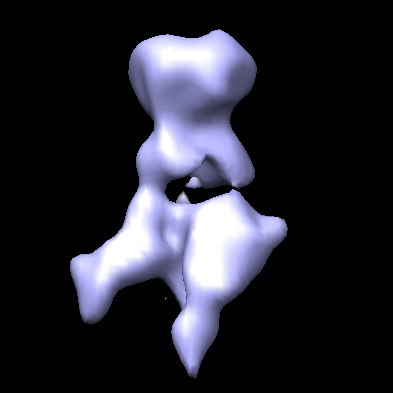} & \includegraphics[width=58pt]{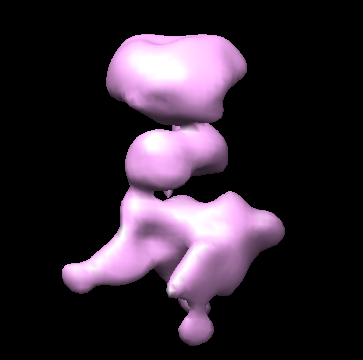} & \includegraphics[width=58pt]{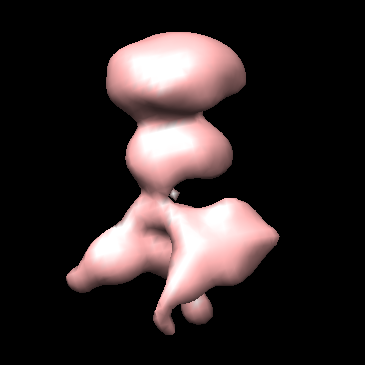} &
\includegraphics[width=58pt]{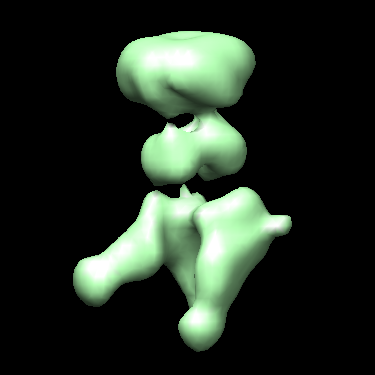} &
\includegraphics[width=58pt]{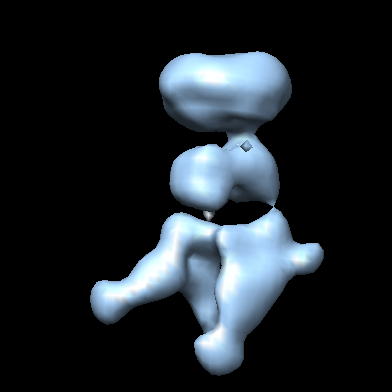} &

\tabularnewline

$\sigma_z = 15$ & \includegraphics[width=58pt]{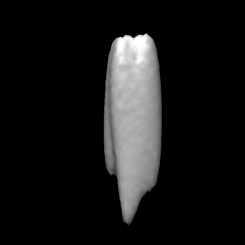} & \includegraphics[width=58pt]{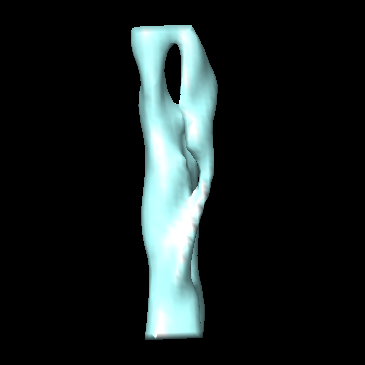} & \includegraphics[width=58pt]{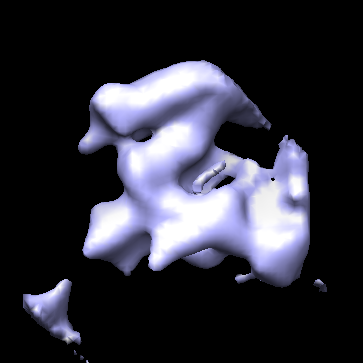} & \includegraphics[width=58pt]{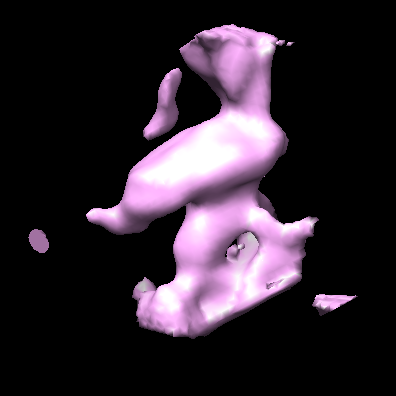} & \includegraphics[width=58pt]{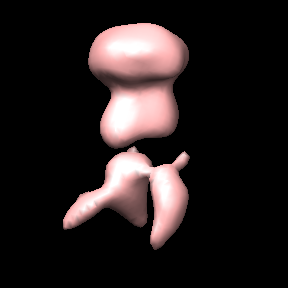} &
\includegraphics[width=58pt]{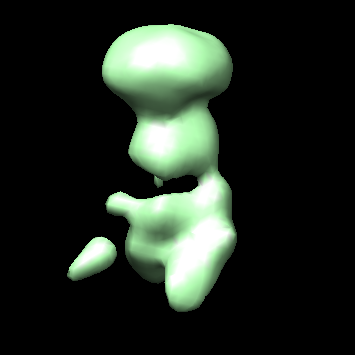} &
\includegraphics[width=58pt]{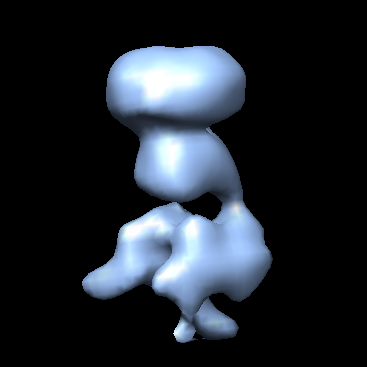} &

\end{tabular}
\caption{Reconstructions obtained with different numbers of views and anisotropy levels $\sigma_z$.
}
\label{nb_view_anis_visual_res}
\end{figure*}

In order to evaluate the robustness of our method to different levels of anisotropy, we performed reconstructions on data generated  with different values of $\sigma_z$. The visual results are given on Figure \ref{nb_view_anis_visual_res}, where we also show the impact of the number of views. When the number of views increases, the reconstruction algorithm is able combine more complementary information, which improves the quality of the reconstruction. As the anisotropy increases, a single views contains less information and the minimum number of views to obtain a satisfying reconstruction also increases. For $\sigma_z = 5$, the minimum number of views is approximately 20, it is about 40 views for $\sigma_z=10$, and for an extremely high anisotropy level of $\sigma_z = 15$, more than 60 views would be necessary to get a correct reconstruction.  

\subsubsection{Robustness to low DOL} 

\begin{figure}
\centering
\begin{tabular}{c@{\hspace{2pt}}c@{\hspace{5pt}}|@{\hspace{5pt}}c@{\hspace{2pt}}c}

\multicolumn{2}{c}{\centering \bf {MP3DR}} & \multicolumn{2}{c}{\centering \bf Our method}\\
\includegraphics[width=80pt]{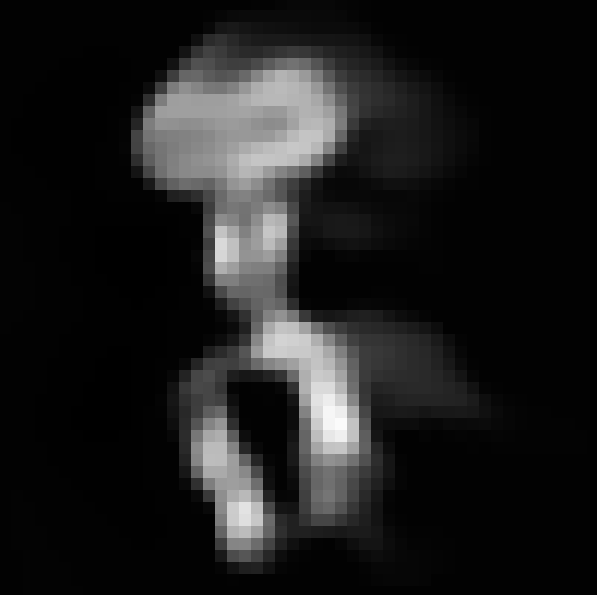} & \includegraphics[width=80pt]{illustrations/illustr_views/recepteurs_AMPA/fortun/partial_labelling/cut.png}&
\includegraphics[width=80pt]{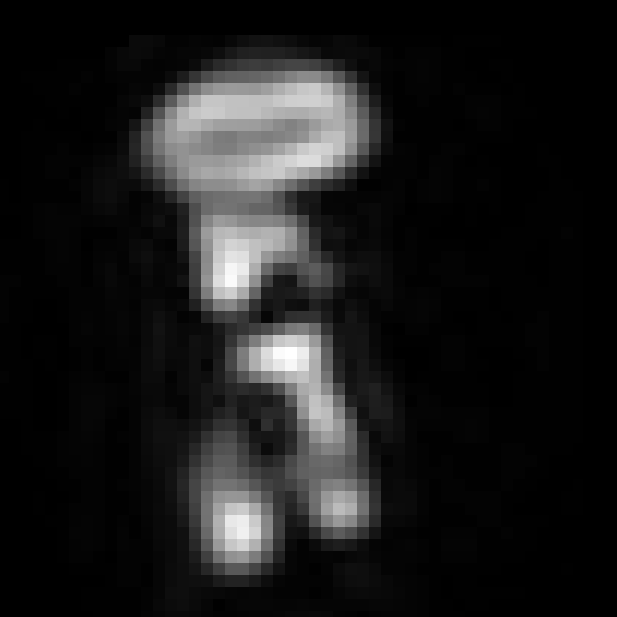} & \includegraphics[width=80pt]{illustrations/illustr_views/recepteurs_AMPA/ours/unknown_angles/partial_labelling/cut.png}\\
\includegraphics[width=80pt]{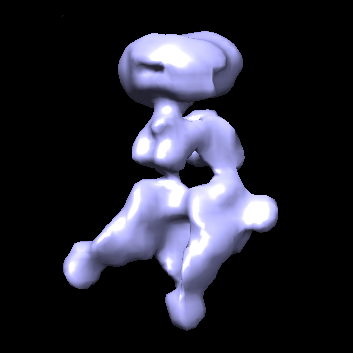} & \includegraphics[width=80pt]{illustrations/illustr_views/recepteurs_AMPA/fortun/partial_labelling/chimera.png}&
\includegraphics[width=80pt]{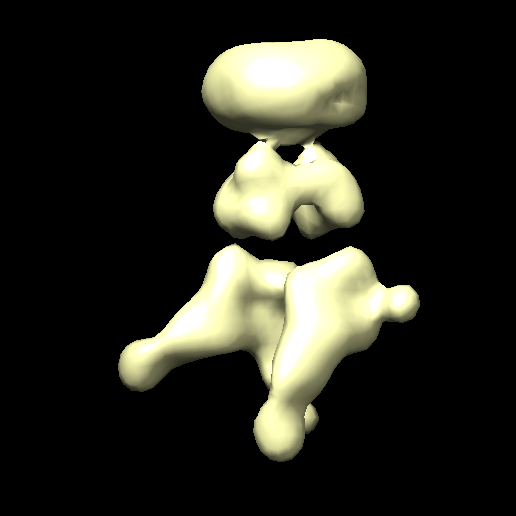} & \includegraphics[width=80pt]{illustrations/illustr_views/recepteurs_AMPA/ours/unknown_angles/partial_labelling/chimera.png}\\
High DOL & Low DOL & High DOL & Low DOL \\
\end{tabular}
\caption{Comparison of the reconstruction of our method and {MP3DR} in case of low and high DOL.}
\label{res_partial_labeling}
\end{figure}

In this section, we focus on the comparison with {MP3DR} and evaluate the impact of the DOL. Let us recall that we simulate the variation of DOL along the structure by subtracting Gaussian spots at random locations. In Figure \ref{res_partial_labeling}, we show reconstructions obtained with our method and {MP3DR} in two different experimental conditions:

\begin{itemize}
\item Low DOL: Default simulation parameters used to simulate the data shown in Figure \ref{fig:visu_views}. It corresponds to a realistic situation.
\item High DOL: {no Gaussian spot is removed from the ground truth image.}
\end{itemize}
 
Our method is able to achieve satisfying reconstructions in both cases, whereas {MP3DR} fails in case of low DOL. This is explained by the sequential nature of {MP3DR}: a first reconstruction with two views is performed, and the other views are then introduced one by one. The flaw of this procedure is that in case of low DOL, the views are too different to achieve a correct pose estimation with only two of them, and it is necessary to combine the information of all the views jointly. Therefore, the initial two-views reconstruction is inaccurate and introduces a detrimental bias that is propagated when the other views are sequentially added. On the contrary, since our approach uses all the views jointly in an unordered way, we are able to use all the information and achieve good reconstruction.

\subsubsection{Sensitivity to hyperparameters}
\label{part:sensity}

\begin{figure*}
\centering
     \begin{tabular}{c@{\hspace{10pt}}c}

     \begin{subfigure}[b]{0.5\textwidth}
     \centering
     \includegraphics[width=\textwidth]{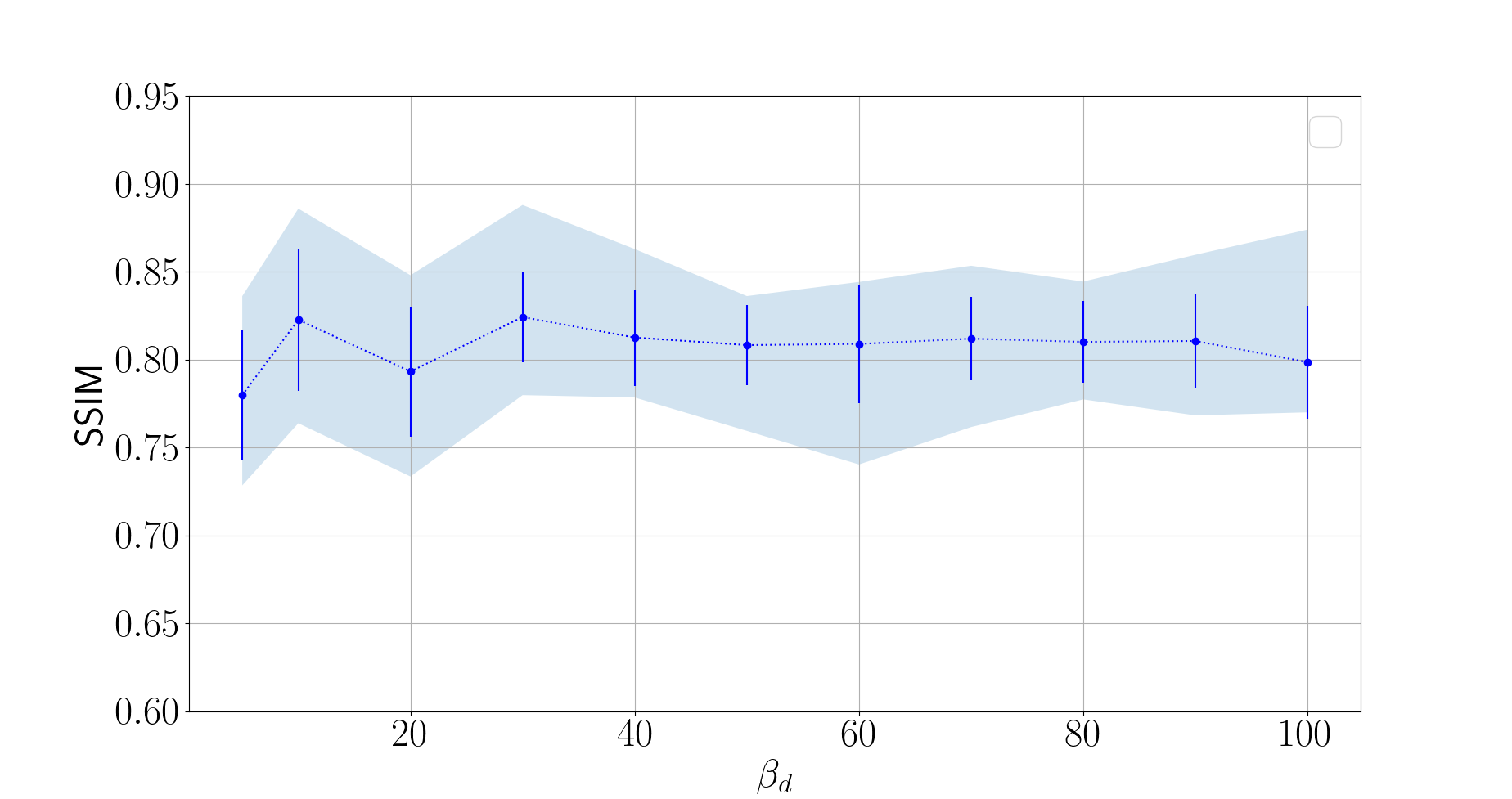}
     \caption{}
     \label{fig:tst_beta_d}
     \end{subfigure}
     &
     \begin{subfigure}[b]{0.5\textwidth}
     \centering
     \includegraphics[width=\textwidth]{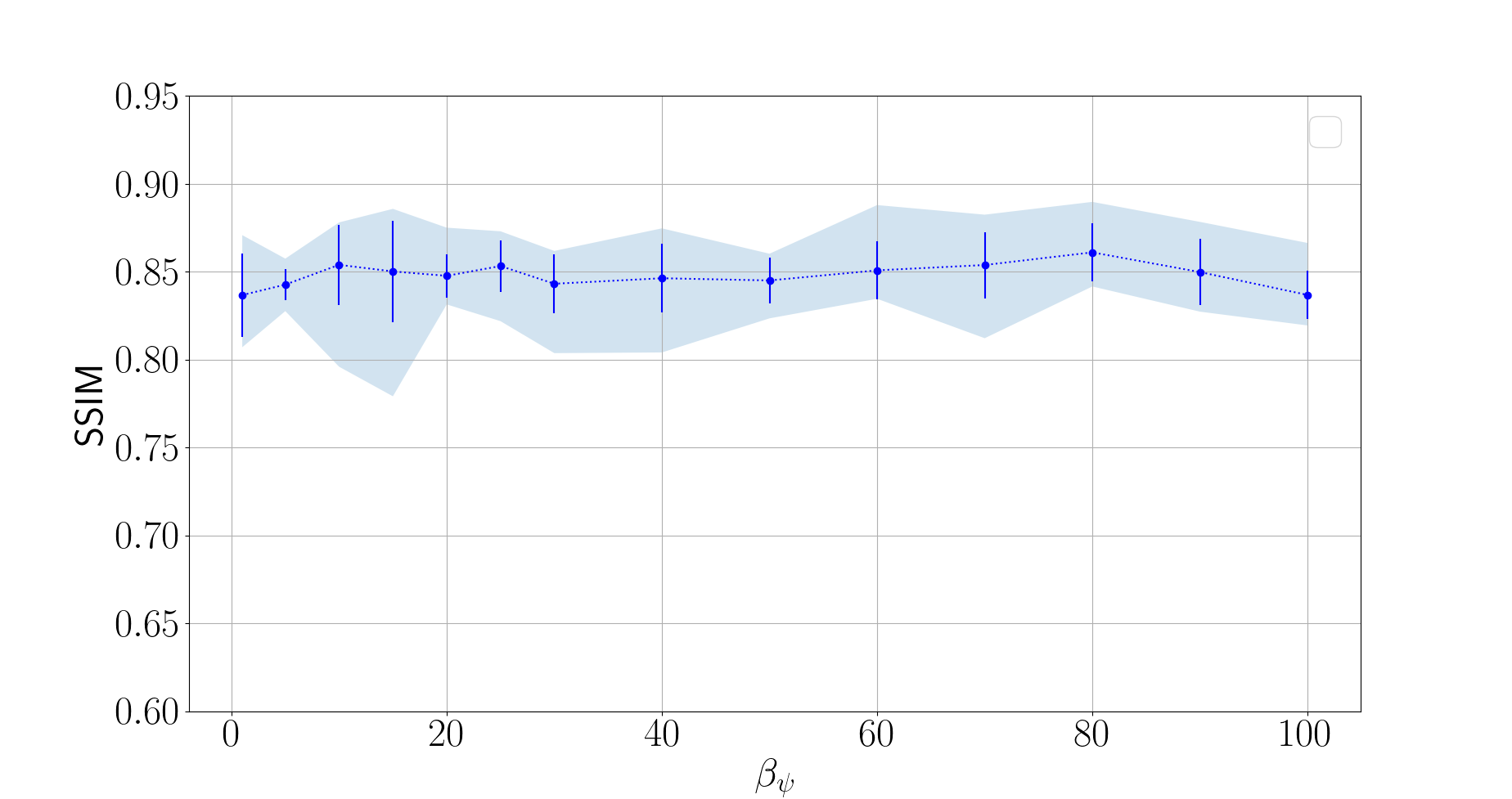}
     \caption{}
	\label{fig:tst_beta_psi}     
     \end{subfigure}
     \\[-5pt]
     \begin{subfigure}[b]{0.5\textwidth}
     	\centering
     	\includegraphics[width=\textwidth]{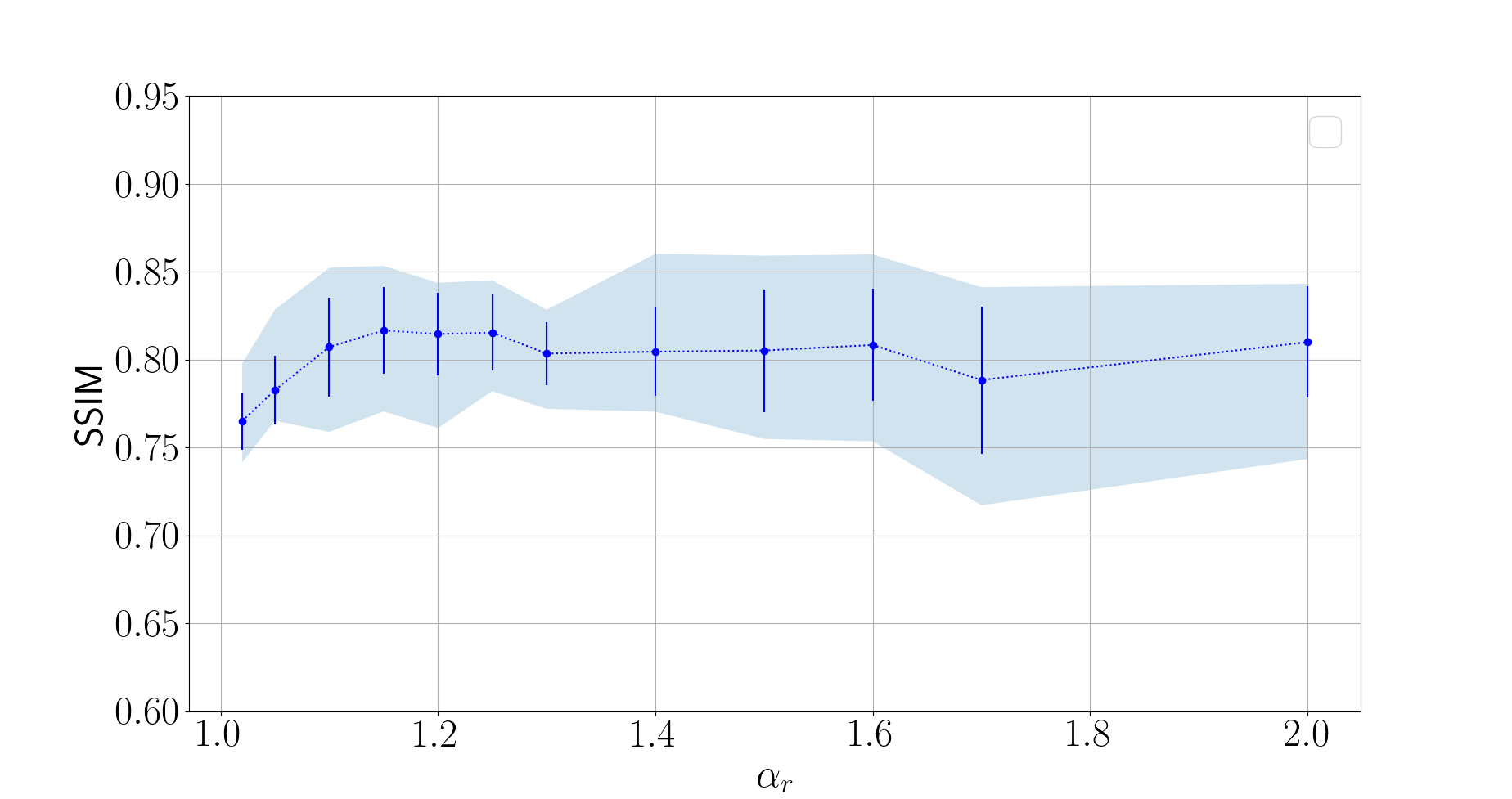}
     	\caption{}
     \label{fig:tst_alpha_r}
     \end{subfigure}
     &
     \begin{subfigure}[b]{0.5\textwidth}
         	\centering
         	\includegraphics[width=\textwidth]{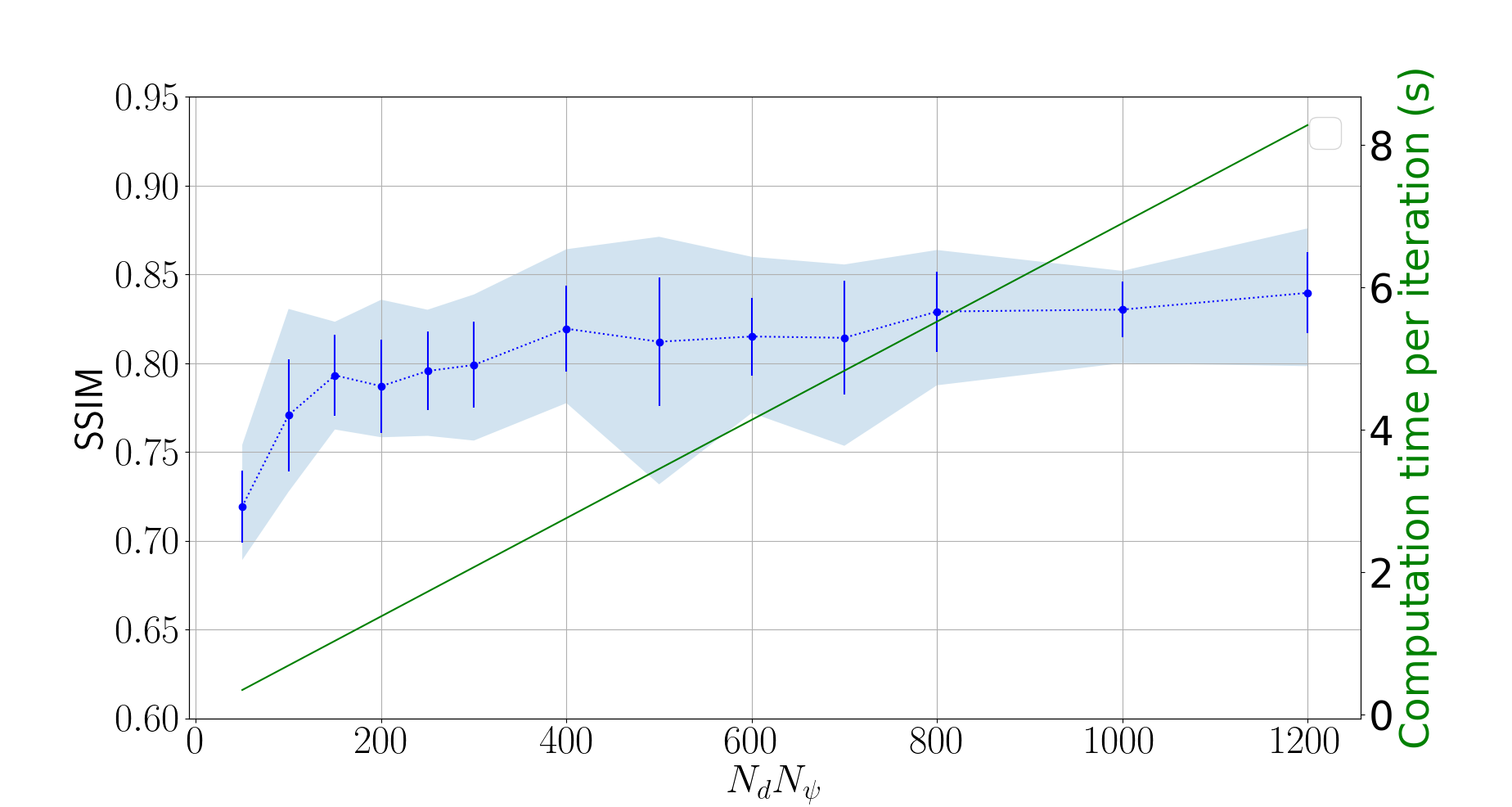}
         \caption{}
     \label{fig:tst_nb_drawn}   
     \end{subfigure}
     
     \end{tabular}
\caption{Influence of the hyperparameters $\beta_d$, $\beta_\psi$, $\alpha_r$ and $N_d N\psi$ on the SSIM and the computation time per iteration. Each point is an average performed on 10 reconstructions on the AMPA receptor. The vertical bars represent the standard deviation on the 10 experiments, and the blue area is the range between the minimum and maximum values.}
\label{fig:tst_hyp_param}
\end{figure*}

A low sensitivity to hyperparameters is crucial to avoid tedious and difficult fine tuning on each new dataset. In this section, we show the robustness of our method to the variation of the following hyperparameters:
\begin{itemize}
\item The ratio $1/\alpha_r$ of the geometric progression at each epoch of the proportion $\alpha$ of uniform component in the distributions $\mathcal{Q}^{l,\psi}$ and $\mathcal{Q}^{l,d}$ \eqref{Q}.
\item The parameters $\beta_d$ and $\beta_{\psi}$ that determine the width of the kernels used for the estimation of the marginal likelihood distributions (see \eqref{kernel2} and \eqref{kernel1}) 
\item The number of orientations sampled at each iteration $N_d$ and $N_{\psi}$. 
\end{itemize}
For each combination of hyperparameters, we performed 10 reconstructions, and report the average SSIM index. 
The default values are: $N_{\psi}=8$, $N_d = 64$, $\alpha_r=1.2$, $\beta_{\psi} = \beta_{d} = 50$, and values are taken in a range around them.    

Figure \ref{fig:tst_hyp_param} illustrates the influence of these hyperparameters on the SSIM between the reconstruction and the ground truth object (the considered particle is AMPA receptors). We observe on Figures  \ref{fig:tst_beta_d} and \ref{fig:tst_beta_psi} that the choice of $\beta_d$ and $\beta_{\psi}$ has a low impact on the final reconstruction. On Figure \ref{fig:tst_alpha_r}, we report the influence of $\alpha_r$ on the SSIM. We observe that low values of $\alpha_r$ (from 1.02 to 1.15) provide slightly lower SSIM, and that the SSIM is relatively stable from $\alpha_r=1.15$ to $\alpha_r=2$. On Figure \ref{fig:tst_nb_drawn}, we report the influence of the number of sampled orientations $N_dN_{\psi}$ in terms of both SSIM and computation time per iteration (We set $N_d = N_{\psi} ^ 2$). We can see that the SSIM increases with the number of sampled orientations. Indeed, if $N_dN_{\psi}$ is too small, the algorithm is likely not to explore the region of the search space which contains the optimal solution. On the other hand, the computation time is proportional to $N_dN_{\psi}$ (see part \ref{part:complex}). Then, choosing $N_d$ and $N_{\psi}$ is a compromise between the quality of the reconstruction and the computation time. 

\label{part:recons_sim}

\subsection{Computation time}
\label{part:complex}

The computation time for one epoch depends on three parameters: the number of views $N$, the number of sampled orientations $N_dN_{\psi}$ and the number of voxels $N_{v}$ in the volume. Each epoch contains as many iterations as the number of views, thus the computation time per epoch is proportional to $N$. 
Most of the computation is devoted to evaluate the energies corresponding to the randomly sampled rotations (lines \ref{rot} to \ref{energy_calc} of Algorithm \ref{algo_summary}). Thus, the computation time per iteration is proportional to $N_dN_{\psi}$. Let us give more detail:

\begin{itemize}
\item In line \ref{rot}, $\hat{f}$ is rotated with orientation $\theta_{i,j}$, then a term by term multiplication is performed. The computation time devoted to the multiplication and to a rotation are $O(N_v)$.
 
\item In line \ref{trans_est}, the phase correlation is computed between $\hat h\mathcal{R}_{\theta_{i,j}}(\hat f)$ and $\hat y_l$ to estimate the translation vector $t_l^{*}(\theta_{i,j})$. The complexity is dominated by the Fourier transform and is $\mathcal O(N_v\log(N_v))$. 

\item In line \ref{energy_calc}, the computation of the energy $E_{i,j}^l$ given the rotated volume $\mathcal{R}_{\theta_{i,j}}$ and the translation vector $t_l^{*}(\theta_{i,j})$ is $O(N_v)$. 

\end{itemize}

{To speed up the computation, we use a GPU implementation of the rotation and the phase correlation, available in CuPy\footnote{https://cupy.dev/} and cuCIM\footnote{https://pypi.org/project/cucim/} packages, respectively}.
The computation time devoted to one epoch $T_{ep}$ is $\mathcal O(N\,N_v\log(N_v)N_dN_{\psi})$. With the default parameters $N=20$, $N_d=64$, $N_{\psi}=8$ and $N_v=50 ^ 3$, a complete reconstruction takes approximately {24 minutes on a Titan Xp graphic card}. Comparatively, {MP3DR} takes 90 minutes for a reconstruction with $N=5$ and the same $N_v = 50 ^3$.

\subsection{Experimental results on real data}

\begin{figure}
\centering
\begin{tabular}{m{110pt}@{\hspace{8pt}}m{110pt}@{\hspace{8pt}}m{110pt}@{\hspace{8pt}}m{110pt}}
	\begin{tabular}{m{31pt}@{\hspace{3pt}}m{74pt}}
    	\includegraphics[width=31pt]{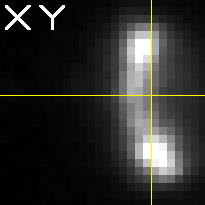} & 
		\includegraphics[width=74pt]{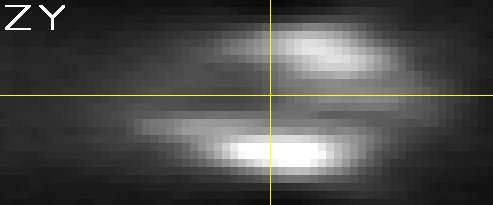}
	\tabularnewline
	\includegraphics[width=31pt]{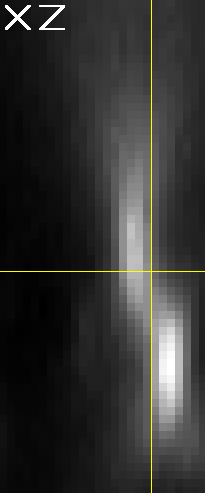} & 
	\includegraphics[width=74pt]{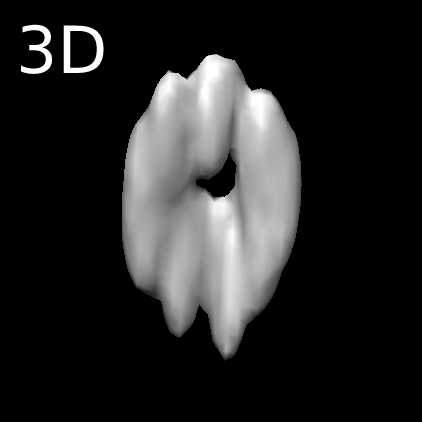} 
	\end{tabular}
&	
	\begin{tabular}{m{31pt}@{\hspace{3pt}}m{74pt}}
    	\includegraphics[width=31pt]{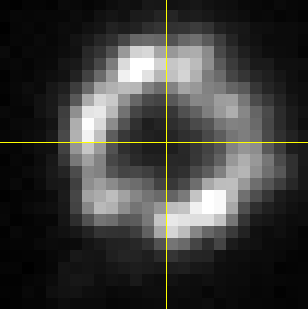} & 
		\includegraphics[width=74pt]{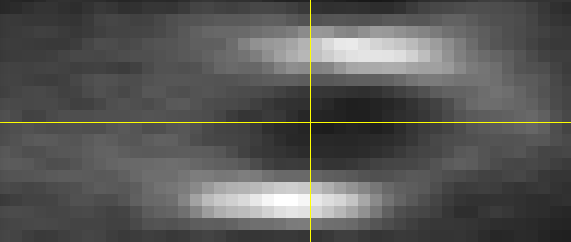}
	\tabularnewline
	\includegraphics[width=31pt]{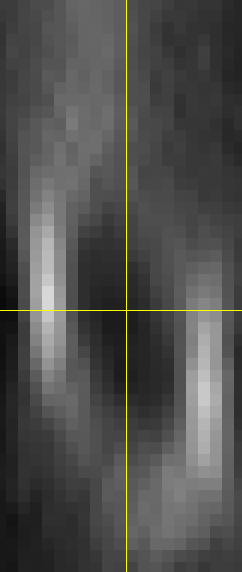} & 
	\includegraphics[width=74pt]{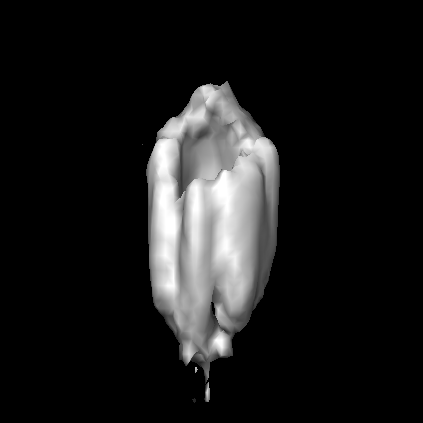} 
	\end{tabular}
&
    \begin{tabular}{m{31pt}@{\hspace{3pt}}m{74pt}}
    	\includegraphics[width=31pt]{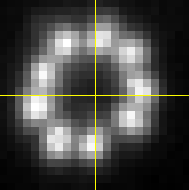} & 
		\includegraphics[width=74pt]{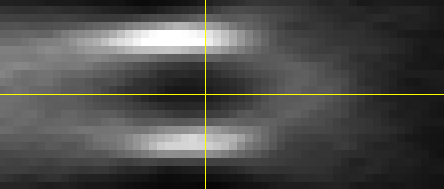}
	\tabularnewline
	\includegraphics[width=31pt]{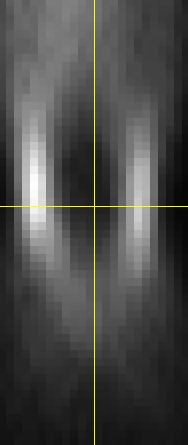} & 
	\includegraphics[width=74pt]{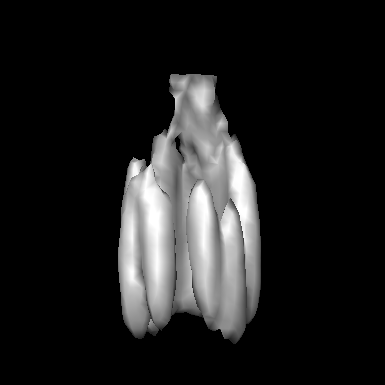} 
	\end{tabular}
 &
 \begin{tabular}{m{31pt}@{\hspace{3pt}}m{74pt}}
    	\includegraphics[width=31pt]{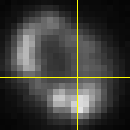} & 
		\includegraphics[width=74pt]{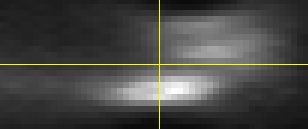}
	\tabularnewline
	\includegraphics[width=31pt]{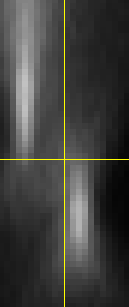} & 
	\includegraphics[width=74pt]{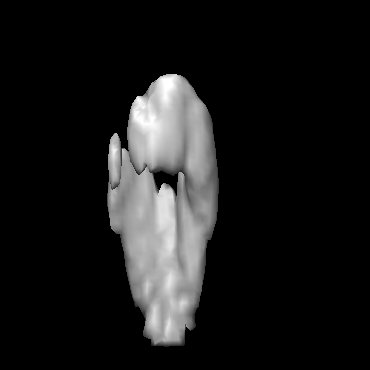} 
	\end{tabular}
	\tabularnewline
	(a) \centering side view & (b) \centering top view & (c) \centering top view & ~~~~~~~~~(d) inclined view \\

\end{tabular}
\caption{Examples of views of our real data set. }\label{fig:views}
\end{figure}

To test our method on real data, we imaged human centrioles from RPE1 cells, an organelle critical for cell division and templating the cilium (PMID: ﻿19914163). As centriole dimensions, about 450nm in length and 250nm in diameter, are below the diffract limit, we imaged them using ultrastructure expansion microscopy (U-ExM). U-ExM is a super-resolution approach that consists in embedding a biological sample into a swellable polymer that will expand 4 times in pure water, allowing to reach a resolution of 50-70nm with a conventional widefield microscope \cite{Hamel21}. After expansion, the biological sample is immuno-labelled with antibodies against Cep164 (2227-1-AP, Proteintech, 1:500), a protein that composes the distal appendages of the centriole \cite{LeGuennec21}, and therefore appears with 9 dots arranged in a ring in super-resolution microscopy \cite{Lau12}. Expanded gels were mounted onto 24 mm coverslips coated with poly-D-lysine (0.1 mg/ml) and imaged with an inverted widefield Leica DM18 microscope. Images were taken with a 63× 1.4 NA oil immersion objective. 3D stacks were acquired with 0.21µm Z-intervals and a 100nm XY pixel size. The PSF was experimentally measured with images of beads. We manually selected 47 views in the acquired volumes. 
The protein Cep164 of the centriole has a ring shape with a ninefold cylindrical symmetry. In Figure \ref{fig:views}, we show 4 examples of views that include 2 top views of the ring, one side view and one inclined view. The resolution along the Z axis is clearly much lower than the resolution on the \XY{} plane. Besides, the non-uniformity of the distribution of fluorophores is observable by comparing the two top views.  The ninefold cylindrical symmetry is also clearly visible on view \ref{fig:views}c.

\begin{figure*}

\centering

\begin{tabular}{m{10pt}m{70pt}@{\hspace{2pt}}m{70pt}@{\hspace{10pt}}m{70pt}@{\hspace{2pt}}m{70pt}@{\hspace{10pt}}m{70pt}@{\hspace{2pt}}m{70pt}}

& \multicolumn{2}{c}{\multirow{2}{*}{\textit{ab-initio} volume}} & \multicolumn{2}{c}{\multirow{2}{*}{Refinement}} & \multicolumn{2}{c}{Refinement with ninefold } \\
& \multicolumn{2}{c}{} 											 & \multicolumn{2}{c}{} 	      & \multicolumn{2}{c}{symmetry constraint} \\

\multirow{2}{*}{\centering \rotatebox{90}{Our method~~~~~}} & \includegraphics[width=70pt]{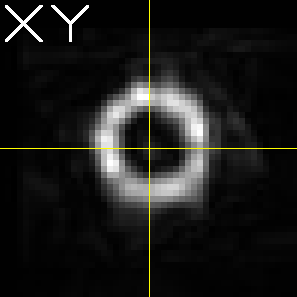} 
& \includegraphics[width=70pt]{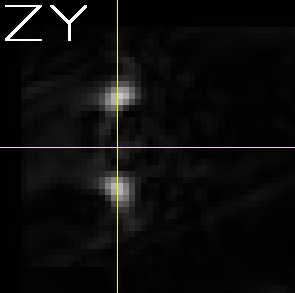} 
& \includegraphics[width=70pt]{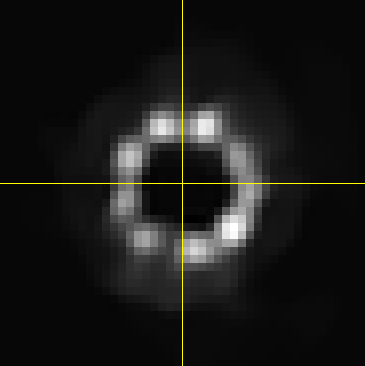} 
& \includegraphics[width=70pt]{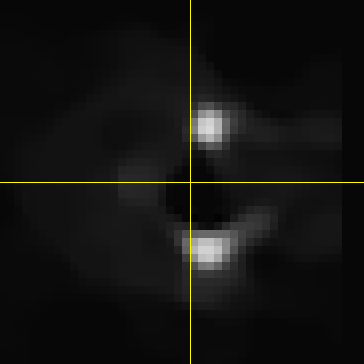} 
& \includegraphics[width=70pt]{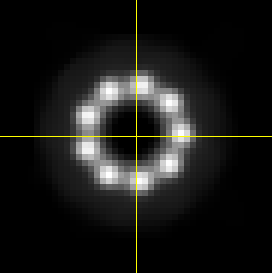} 
& \includegraphics[width=70pt]{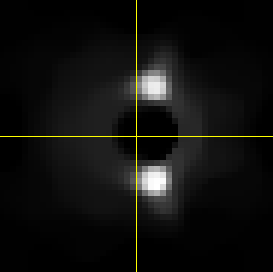} \\[-2pt]

& \includegraphics[width=70pt]{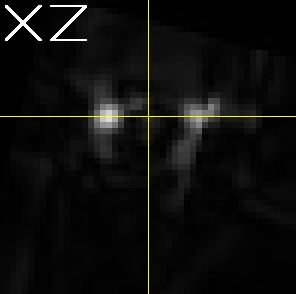} 
& \includegraphics[width=70pt]{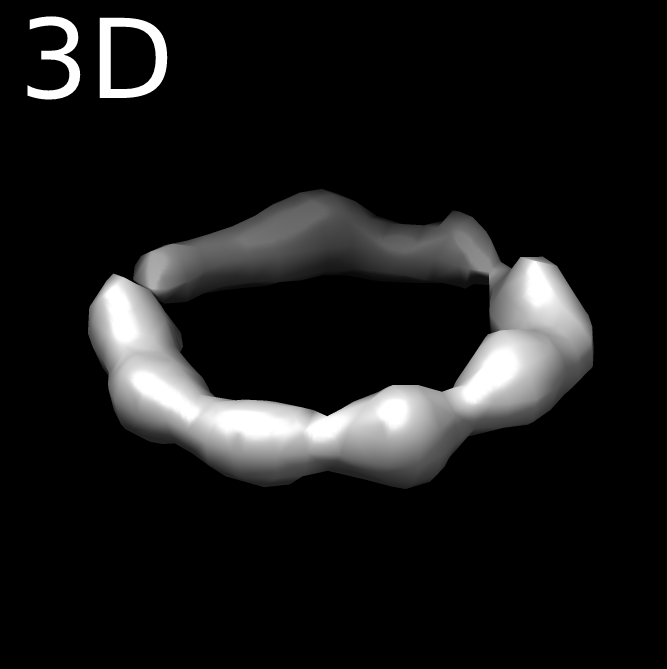} 
& \includegraphics[width=70pt]{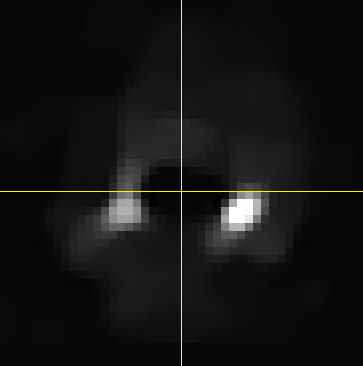} 
& \includegraphics[width=70pt]{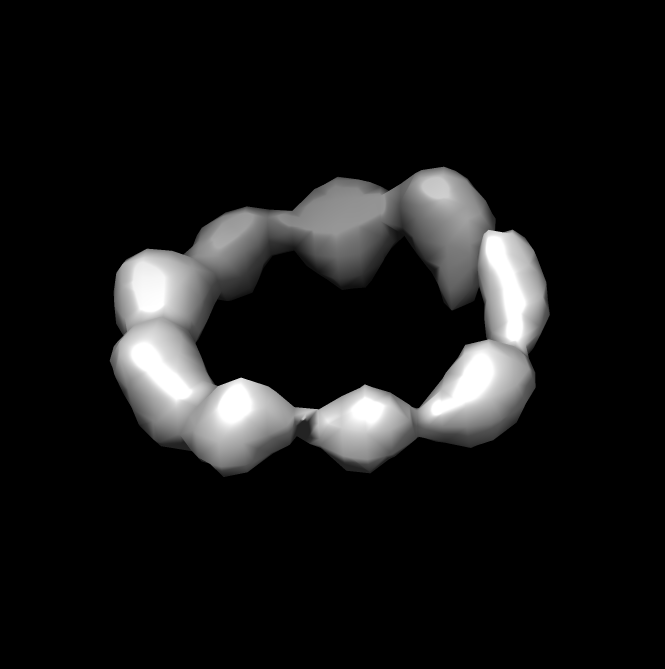} 
& \includegraphics[width=70pt]{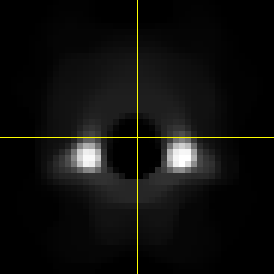} 
& \includegraphics[width=70pt]{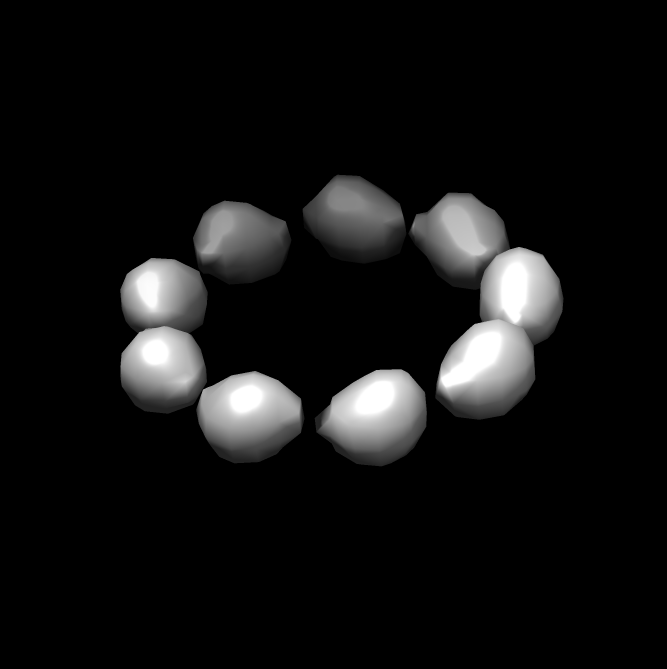}    \\

\\[-5pt]

\multirow{2}{*}{\centering \rotatebox{90}{Cryo-RANSAC~~~}} & \includegraphics[width=70pt]{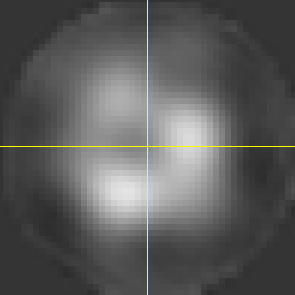} 
& \includegraphics[width=70pt]{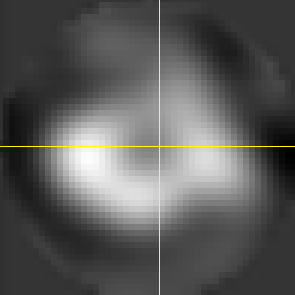} 
& \includegraphics[width=70pt]{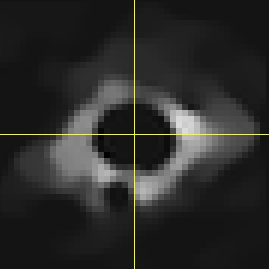} 
& \includegraphics[width=70pt]{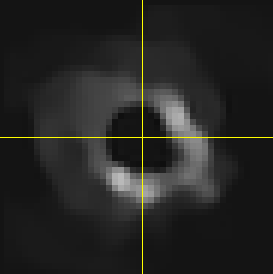} 
& \includegraphics[width=70pt]{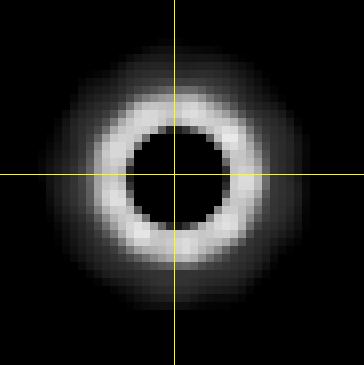} 
& \includegraphics[width=70pt]{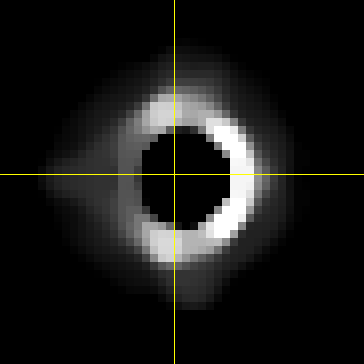} \\[-2pt]

& \includegraphics[width=70pt]{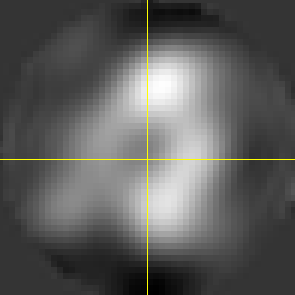} 
& \includegraphics[width=70pt]{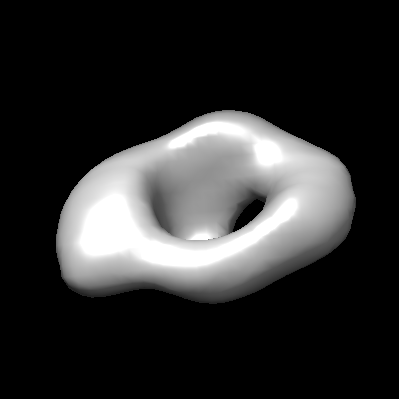} 
& \includegraphics[width=70pt]{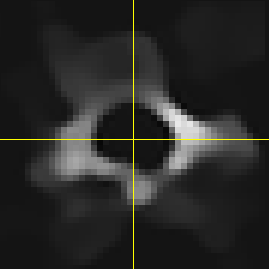} 
& \includegraphics[width=70pt]{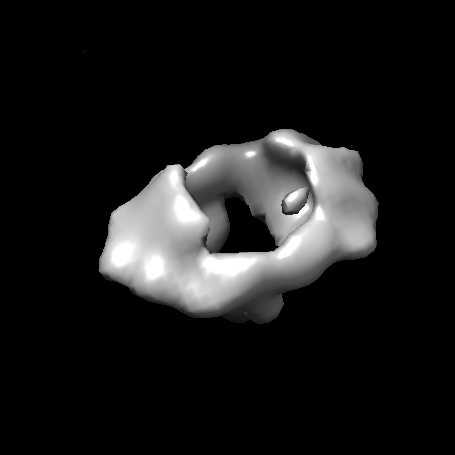} 
& \includegraphics[width=70pt]{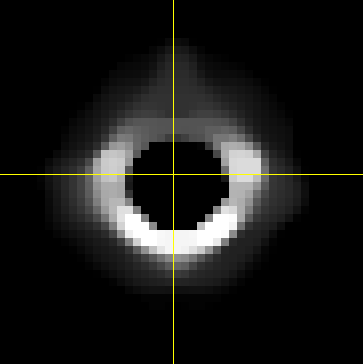} & \includegraphics[width=70pt]{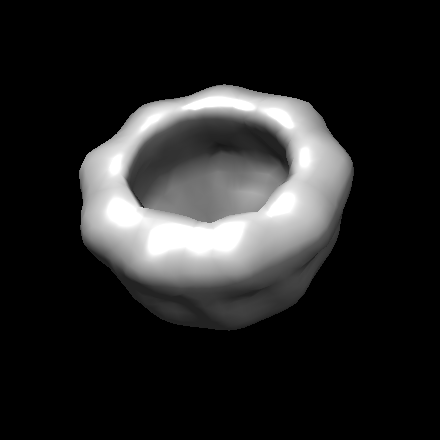} \\

\end{tabular}
\caption{Reconstruction results obtained on a real dataset of centriole, with our method and cryo-RANSAC.}
\label{fig:ourreconsreal}
\end{figure*}

In Figure \ref{fig:ourreconsreal}, we show the reconstruction obtained with this dataset and the same hyperparameters as the ones presented in Section \ref{part:sensity}. We compare our result with the one of cryo-RANSAC. In order to get a high-accuracy result, the {\it ab initio} reconstructions are refined using an alternated optimization scheme described in \cite{Fortun18}. In addition to the basic refinement, we also show reconstructions obtained with an additional nine-fold symmetry constraint.

 The reconstructed shape found by our method is a ring that is consistent with the input views. 
Indeed, the diameter of the ring observable on the \XY{} plane is similar to the diameter of the top views \ref{fig:views}b and \ref{fig:views}c	. 
 Besides, the thickness of the ring, observable on the XZ and ZY cuts is similar to what we observe on the side view \ref{fig:views}a. The ninefold symmetry is fuzzy in the \textit{ab-initio} reconstruction, but sufficient to allow for a clearly visible symmetry in the refined volume. If we impose the ninefold symmetry in the refinement, the ring has a perfectly regular symmetry. The cryo-RANSAC reconstruction is not able to retrieve the same details, since the ring shape is not correctly reconstructed, and the ninefold symmetry is not visible.

\section{Conclusion}
We proposed a { reference-free} SPR approach that jointly estimates the poses and the reconstruction, to address the limitations of resolution anisotropy and non-uniform DOL in fluorescence microscopy. Our approach is based on a reformulation of the original joint optimization problem in a multilevel scheme. We developed fast approximate solvers at each level to end up with a computationally efficient method able to estimate the particle poses and the reconstruction. We demonstrated experimentally that our method yields more accurate reconstructions than the standard methods, on both simulated and real data. Besides, our method is fast, it is robust to low DOL, it has a very low sensitivity to hyperparameters, and it does not require any particles classification step. These features are crucial for practical applications of the method on a variety of real data.


\section{Acknowledgement}
This work was supported by the French National Research Agency (ANR) through the SP-Fluo project (ANR-20-CE45-0007).

The authors would like to acknowledge the High Performance Computing Center of the University of Strasbourg for supporting this work by providing scientific support and access to computing resources. Part of the computing resources were funded by the Equipex Equip@Meso project (Programme Investissements d'Avenir) and the CPER Alsacalcul/Big Data.

This work is supported by the Swiss National Science Foundation (SNSF) grant PP00P3{\_}187198 and the European Research Council (ERC) StG 715289 to P. Guichard and the SNSF grant 310030{\_}205087 to P. Guichard and V. Hamel.

\bibliographystyle{plain}
\bibliography{biblio2}

\section{Appendix}
\label{grad_comp}

Here we derive the computation of the gradient of the energy $E_l$ with respect to the volume $\hat f$ used in the SGD update equation \eqref{maj}. We have
\begin{eqnarray}
E_l(f, \theta_l, t_l) &=& \|\hat y_l - \hat h  \rho_{t_l}  \mathcal{R}_{\theta_l}(\hat f)  \|^2 \\ 
&=& \int[\hat{y}_l(\omega) - \hat{h}(\omega)\rho_{t_l}(\omega)\hat{f}(R_{\theta_l}^ {T}\omega)] ^ 2 d\omega.  \label{integral}
\end{eqnarray}

Let us consider a particular frequency $\omega_0$ and compute the derivative $\frac{\partial E_l}{\partial \hat{f}}(\omega_0)$.
In the integral \eqref{integral}, only the term that verifies $R_{\theta_l}^T\omega = \omega_0$ depends on $\hat{f}(\omega_0)$: 
\begin{equation}
 \frac{\partial \hat f(R_{\theta_l}^T\omega)}{\partial \hat f}(\omega_0) = \begin{cases}
1 &\text{if $R_{\theta_l}^T\omega = \omega_0$}\\
0 &\text{otherwise}
\end{cases}.
\end{equation}
Then we have 
\begin{small}
\begin{eqnarray}
    \frac{\partial E_l}{\partial \hat f}(\omega_0) = 2 \left [\rho_{R_{\theta_l}t_l}(\omega_0)\hat h(R_{\theta_l}\omega_0)\hat f(\omega_0) - \hat y_l(R_{\theta_l}\omega_0) \right] \\
    \rho_{R_{\theta_l}t_l}(\omega_0) \hat h(R_{\theta_l}\omega_0)\nonumber \\ 
    = 2 \left [\rho_{R_{\theta_l}t_l}(\omega_0)\mathcal{R}_{-\theta_l}(\hat h)(\omega_0)\hat f(\omega_0) - \mathcal{R}_{-\theta_l}(\hat y_l)(\omega_0) \right]\\
    \rho_{R_{\theta_l}t_l}(\omega_0) \mathcal{R}_{-\theta_l}(\hat h)(\omega_0).\nonumber
\end{eqnarray}
\end{small}
Finally, the gradient can be written
\begin{equation}
\begin{multlined}
    \nabla_{\hat f} E_l(f, {\theta^*_l}, {t^*_l})  \\
    = 2\rho_{R_{\theta_l}t_l}\mathcal{R}_{-\theta_l}(\hat h) \left [\rho_{R_{\theta_l}t_l}\mathcal{R}_{-\theta_l}(\hat h) \hat f - \mathcal{R}_{-\theta_l}(\hat{y}_l)) \right].
\end{multlined}
\end{equation}

\end{document}